\documentclass[aps,prl,twocolumn,showpacs,notitlepage,floatfix,superscriptaddress]{revtex4-1}

\usepackage{amsmath}
\usepackage{amssymb}
\usepackage{amsfonts}
\usepackage{mathtools}
\usepackage{bm}             

\usepackage{graphicx}
\usepackage{epstopdf}
\usepackage{subfigure}

\usepackage[dvipsnames]{xcolor}

\usepackage[breaklinks=true,colorlinks=true,linkcolor=RubineRed,citecolor=Blue,filecolor=Blue,urlcolor=Blue,pdfstartview=FitH]{hyperref}
\usepackage{breakcites}

\usepackage{dcolumn}         
\usepackage{multirow}        
\usepackage{braket}          
\usepackage[american]{babel}
\usepackage{textcomp}        
\usepackage{comment}
\usepackage{orcidlink}       
\usepackage[normalem]{ulem}

\makeatletter
\def\maketitle{
\@author@finish
\title@column\titleblock@produce
\suppressfloats[t]}
\makeatother

\begin{document}
\title{Circulation Statistics and Migdal Area Rule Beyond the Kibble-Zurek Mechanism in a Newborn Bose-Einstein Condensate}

\author{Matteo~Massaro\orcidlink{0009-0004-0935-8022}}
\email{matteo.massaro@uni.lu}
\affiliation{Department  of  Physics  and  Materials  Science,  University  of  Luxembourg,  L-1511  Luxembourg,  Luxembourg}

\author{Seong-Ho Shinn\orcidlink{0000-0002-2041-5292}}
\email{seongho.shin@uni.lu}
\affiliation{%
  Department of Physics and Materials Science, University of Luxembourg, L-1511 Luxembourg, Luxembourg}
\author{Mithun Thudiyangal\orcidlink{0000-0003-4341-6439}}
\affiliation{Center for Quantum Technologies and Complex Systems (CQTCS), Christ University, Bengaluru, Karnataka 560029, India}
\affiliation{Department of Physics and Electronics, Christ University, Bengaluru, Karnataka 560029, India}

\author{Adolfo del Campo\orcidlink{0000-0003-2219-2851}}
\affiliation{Department  of  Physics  and  Materials  Science,  University  of  Luxembourg,  L-1511  Luxembourg,  Luxembourg}
\affiliation{Donostia International Physics Center,  E-20018 San Sebasti\'an, Spain}

\begin{abstract}
The Kibble-Zurek mechanism (KZM) predicts that a newly formed superfluid prepared by a finite-time thermal quench is populated with vortices. The universality of vortex number statistics, beyond KZM, enables the characterization of circulation statistics within any region of area $A$ enclosed by a loop $C$. Migdal's minimal 
area rule of classical turbulence predicts that the probability density function of circulation around a closed contour is independent of the contour’s shape. We verify the Migdal area rule for small loops with respect to the distance between the vortex and antivortex pairs and further characterize its universal breakdown for bigger loops. We further uncovered the nonequilibrium universality dictated by the KZM dynamics, which results in power-law scalings of the moments of the circulation statistics as a function of the quench time.
\end{abstract}
\maketitle

The study of circulation statistics plays a key role in understanding turbulence in both classical and quantum fluids. Yet, in two- and three-dimensional quantum turbulence, it is characterized by non-universal behavior. In this context, Migdal introduced an area rule for classical inviscid fluids \cite{Migdal94,migdal2019universal,migdal2019exact}, according to which the circulation statistics associated with a minimal surface area enclosed by a loop depends only on the loop area and is insensitive to the shape of the loop.
The circulation $\Gamma$ of the fluid within the area $A$ satisfies 
\begin{equation}
\left\langle 
\Gamma^p 
\right\rangle 
\propto 
\left\vert 
A 
\right\vert^{\alpha \left( p \right) p}
, 
\label{Migdal_circulation}
\end{equation}
with $\alpha \left( p \right) = 2/3$ for small $p$ and $\alpha \left( 10 \right) \simeq 0.58$ \cite{migdal2019universal}, 
where $\left\langle x \right\rangle$ is the average of $x$ and $\left\vert A \right\vert$ is the area of $A$. 
Migdal's work has inspired further theoretical \cite{Iyer19,Apolinario20,Iyer21,Muller21,Xie2025} and experimental studies \cite{Zhu23}.
Within mean-field theory, connections between classical and quantum descriptions are anticipated. Specifically, the dissipative Gross-Pitaevskii (or nonlinear Schr\"odinger) equation can be mapped onto the Navier-Stokes equation with an additional quantum potential term \cite{Tsatsos16}. Although a formal proof of the area rule remains to be established, both in the classical and quantum domains, it is natural to explore its implications in superfluids \cite{Muller21, Polanco21}.

The spontaneous breaking of $U(1)$ symmetry in finite time leads to the formation of vortices \cite{Weiler08,DRP11,Chesler15,Zeng23,Thudiyangal24}. The Kibble-Zurek mechanism (KZM) dictates that the average density of vortices scales as a universal power law of the driving rate at which the phase transition is crossed  \cite{Kibble76a,Kibble76b,Zurek96a,Zurek96b,DZ14}. Specifically, consider a continuous phase transition as a function of a control parameter $\lambda$ with a critical point at $\lambda_c$. As a function of the dimensionless distance to the critical point $\varepsilon=(\lambda_c-\lambda)/\lambda_c$, the equilibrium correlation length scales as $\xi=\xi_0/|\varepsilon|^\nu$, while the relaxation time obeys $\tau=\tau_0/|\varepsilon|^{z\nu}$, where $\nu$ and $z$ are critical exponents associated with the universality class of the system. For a linearized quench satisfying $\varepsilon=t/\tau_Q$, the KZM dictates that the nonequilibrium correlation length scales universally as a function of the quench time $\tau_Q$ according to $\hat{\xi}=\xi_0(\tau_Q/\tau_0)^{\nu/(1+z\nu)}$.
It follows that the density of spontaneously formed vortices in the newborn superfluid scales as $n\propto\hat{\xi}^{-2}$, in two spatial dimensions. 
One may wonder whether the newborn superfluid exhibits spontaneous quantum turbulence. This possibility is supported by recent numerical simulations \cite{Yang2025,shinn2025}. In particular, the Kolmogorov scaling \cite{Tsubota13,Tsatsos16}, characterizing the dependence of the incompressible kinetic energy on the wavevector, has been reported in two spatial dimensions, confirming that KZM leads to spontaneous quantum turbulence in a newborn Bose-Einstein condensate \cite{shinn2025}.


In this Letter, we investigate the spontaneous quantum turbulence (SQT) resulting from Bose-Einstein condensation in finite time and establish the universality of the circulation statistics within an area $A$ enclosed by a loop $C$. In doing so, we confirm the validity of Midgal's  
area rule in describing a newborn superfluid and further characterize its breakdown. We further uncover the nonequilibrium universality imposed by the Kibble-Zurek dynamics on the area rule, leading to scaling laws of the circulation statistics moments as a function of the rate driving the superfluid formation.

\begin{figure*}[hptb]
    \centering
    \includegraphics[width=1.0\linewidth]{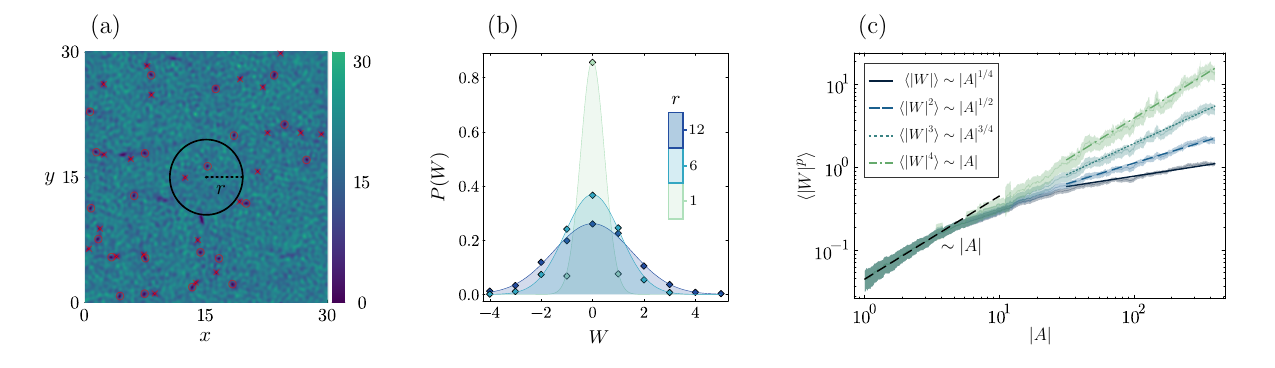}
    \caption{\textbf{Typical condensate density at equilibration time and corresponding circulation statistics.} 
    Panel (a) depicts the condensate density \(|\Psi_{\mathcal{C}}(\mathbf{r},t_{eq})|^{2}\) at equilibration time \(t_{eq}\) after a quench of duration \(\tau_{Q}=20\), with vortices of topological charge \(w=+1\) marked by crosses and \(w=-1\) by circles. Higher‐charge vortices \(|w|>1\), being energetically unfavorable, decay rapidly and are never observed in our simulations.
   Panel (b) shows the 
   PMF 
   of the net charge \(W\) within circular loops of radius \(r\), each averaged over \(\mathcal{R}=1000\) independent realizations. 
   Panel (c) displays the corresponding moments of \(|W|\) as a function of the area $|A|=\pi r^2$, confirming the scaling predicted by Eq.(\ref{eq:W_every_powers_A}). Shaded error bands denote $95\%$ confidence intervals.
}
\label{fig:density_and_phase_psi_with_marked_vortices_and_KZ_scaling}
\end{figure*}

{\it Circulation statistics beyond the KZM.} 
Early KZM studies pointed out that the spontaneous quantized current in a superconducting or superfluid ring after a thermal quench can be described as the result of a random walk \cite{Zurek96c}. This prediction has stimulated a series of confounding experiments \cite{Monaco02,Monaco03,Monaco06,Monaco09} and theoretical efforts \cite{Das12,Zurek13,Sonner15,Zeng:2019yhi}.
Recent generalizations of the KZM have further uncovered the universality of the number distribution of topological defects \cite{delcampo18,GomezRuiz20,delCampo21}, as well as their spatial correlations \cite{delcampo22,Thudiyangal24}, beyond the KZM. In a newborn superfluid prepared in finite time, the circulation statistics can thus be predicted by such generalizations of the KZM framework. Consider the 
probability mass function (PMF) 

\begin{equation}
P(\Gamma)=\left\langle \delta\left(\Gamma-
\oint_{C} v(\mathbf{r},t)\cdot d\ell\right)\right\rangle. 
\end{equation}
Given the quantization condition $\Gamma= \left( 2 \pi \hbar / m \right) W$ with $m$ being the mass of the particle in the ultracold gas,
$
P(\Gamma)= \left( m / 2\pi\hbar \right) P( W )
$
where 
$P( W )$ is the 
PMF 
of the imbalance  $W$ between the number of vortices and antivortices within the area enclosed by $C$, 
\begin{eqnarray}
W \coloneqq n_+ - n_-.
\end{eqnarray}
This assumes that the topological charge $w$ of each vortex is either $1$ or $-1$, which is consistent with the instability of defects with higher charge values. Here, $n_{\pm}$ denotes the number of vortices with topological charge $\pm 1$. The average of this quantity is expected to be zero in the absence of any deterministic source of angular momentum, i.e., $\langle W\rangle = \langle\Gamma\rangle = 0$.  

In a related context, Zurek estimated the variance of $W$ in the limit of a large total number of vortices $n = n_+ + n_-$ using a random walk argument, which yields $\sqrt{\langle W^2\rangle} \propto \sqrt{n}$ \cite{Zurek96c,Zurek13}. This prediction has been verified in numerical studies involving $U(1)$ symmetry breaking \cite{Das12,Nigmatullin16,Xia2020} as well as experiments \cite{Lin14}.  An alternative derivation, based on the random phase of the superfluid wavefunction, gives $\sqrt{\langle W^2 \rangle} \propto \sqrt{|C| / \hat{\xi}}$, where $\hat{\xi}$ is the Kibble-Zurek correlation length.  

Studies of the spatial statistics of vortices in scenarios involving spontaneous $U(1)$ symmetry breaking have led to the introduction of the PPP-KZM model, according to which vortices are distributed via a spatially homogeneous Poisson Point Process (PPP) with KZM density \cite{delcampo22,Thudiyangal24}. 
It follows that
\begin{eqnarray}
&&
\langle \Gamma^2 \rangle 
\propto 
\left\vert 
A 
\right\vert 
/ 
\hat{\xi}^2
, 
\quad 
\textrm{ from the PPP-KZM,}
\nonumber\\
&&
\langle \Gamma^2 \rangle 
\propto 
\left\vert 
C 
\right\vert 
/ 
\hat{\xi}
, 
\quad 
\textrm{ from the random phase.}
\end{eqnarray}
Experiments to date on multiferroic hexagonal manganite show that 
$
\langle \Gamma^2 \rangle 
\propto 
\left\vert 
A 
\right\vert
$ for 
$\left\vert 
A 
\right\vert < \left\vert 
A_{c_1} 
\right\vert$, and 
$
\langle \Gamma^2 \rangle 
\propto 
\left\vert 
C 
\right\vert
$ for 
$\left\vert 
A 
\right\vert > \left\vert 
A_{c_2} 
\right\vert$, with $\left\vert A_{c_1} \right\vert < \left\vert A_{c_2} \right\vert$ \cite{Lin14}.

{\it Kibble-Zurek dynamics of the condensate.} 
We simulate the condensate formation using the two-dimensional (2D) stochastic projected Gross--Pitaevskii equation \cite{Gardiner_2003,blakie2008dynamics,Rooney2012,Rooney2014,Bradley2015,McDonald2020}. The main idea behind this framework is to divide the modes of the system into two regions based on an energy cutoff: the low-energy, highly populated modes constitute the so-called coherent region ($\mathcal{C}$), while the remaining higher-energy states form the incoherent region ($\mathcal{I}$), which is characterized by a low occupation number and acts as a thermal reservoir.
The dynamics of the condensate band is captured by a field $\Psi_{\mathcal{C}}(\mathbf{r},t)$, which evolves according to
\begin{equation}\label{2D_dimensionless_SPGPE_main}
    d{\Psi_{\mathcal{C}}} = \mathcal{P}_{\mathcal{C}} \left[ -(i + \gamma) \left( H_{sp} + g |\Psi_{\mathcal{C}}|^{2} -\mu\right) \Psi_{\mathcal{C}} \, dt + d \eta \right].   
\end{equation}
Here, $\mathcal{P}_{\mathcal{C}}$ is the projection operator onto the $\mathcal{C}$ subspace. 
Furthermore, $H_{sp} = - \left( 1 / 2 \right) \nabla^{2} + V(\mathbf{r})$ is the single-particle Hamiltonian,  $\gamma$ denotes the dissipation rate, and $d\eta$ is a complex Gaussian noise increment satisfying the fluctuations dissipation theorem $\langle d\eta(\mathbf{r},t) d\eta^{*}(\mathbf{r}^{\prime},t) \rangle=2\gamma T \delta_{\mathcal{C}}(\mathbf{r},\mathbf{r}^{\prime}) dt$, with $T$ denoting the temperature of the thermal cloud and $\delta_{\mathcal{C}}$ the Dirac delta in the $\mathcal{C}$ region.
Throughout this work, we consider a periodic homogeneous system by setting the trapping potential $V(\mathbf{r},t)=0$. 
Further details of Eq.~(\ref{2D_dimensionless_SPGPE_main}) along with its numerical implementation can be found in \cite{shinn2025}.

The BEC transition is driven by tuning the chemical potential via a linearized quench $\mu(t) = \mu_i + \left( \mu_f - \mu_i \right)t/\tau_Q$, where $\mu_i$ and $\mu_f$ are the initial and final values.
The crossing of the critical point $\mu_c$ at a finite rate $\tau_Q$  
yields a Bose-Einstein condensate seeded with vortices, whose average density is predicted by the KZM to scale with the quench rate as $ n \propto \tau_{Q}^{-2\nu/(1+z\nu)}$. Specifically, in the mean-field regime, where $\nu = 1/2$ and $z = 2$, $ n \propto \tau_{Q}^{-1/2}$ \cite{Thudiyangal24, shinn2025}.

In the following, we address physics beyond the conventional scope of the KZM by studying circulation statistics of the superfluid velocity and explore the validity in the quantum domain of Migdal's area rule of classical turbulence.
To measure the circulation around a given loop in the newborn condensate, we detect the vortices trapped inside the loop at equilibration time, which marks the switch in the condensate density growth from exponential to linear and is proportional to the Kibble–Zurek freeze‑out time \cite{Chesler15}. 
We then determine the corresponding net topological charge by subtracting the number of vortices and anti-vortices, since multiply charged vortices (\(|w|>1\)) are energetically suppressed and decay rapidly.
 A visual representation of this procedure is given in Fig. \ref{fig:density_and_phase_psi_with_marked_vortices_and_KZ_scaling} (a). 

{\it Circulation statistics via the vortex pair model.} 
Inspired by Zurek’s insight \cite{Zurek13}, we model the spatial distribution of vortices at equilibration time as an ensemble of vortex–antivortex pairs, represented by segments of length $\hat{\xi}$ connecting opposite topological charges. The number of pairs, $N$, fluctuates across different system realizations and can be described by $\mathcal{N}$ independent Bernoulli trials with probability of success $p_v$ \cite{GomezRuiz20}. Specifically, $\mathcal{N} = A_{\rm tot}/(2\,f\,\hat{\xi}^2)$ is half the number of possible vortex nucleation sites in a system of total area $A_{\rm tot}$, with $f$ a geometric fudge factor \cite{Laguna97,Yates98,GomezRuiz20,Balducci23}, and $p_v$ the vortex formation probability at each site. Hence, the distribution is given by $P(N = n) = \binom{\mathcal{N}}{n}\,p_v^n\,(1 - p_v)^{\mathcal{N}-n}$.

With this model in mind, the net winding number $W$ enclosed by a loop $C$ is determined by the vortex-antivortex segments that intersect its contour, i.e., with one endpoint inside $C$ and the other outside. We denote by $p_{C}$ the probability that a given pair crosses $C$, assuming uniform, independent placement and random orientation. 
Each crossing adds $+1$ or $-1$ to the net charge with equal probability.
We can thus write $W=\sum_{i=1}^{\mathcal{N}} z_{i}$, where $z_{i}$ are i.i.d. random variables taking the value $0$ with probability $(1-p_{v}p_{C})$ and $\pm 1$ with probability $p_{v}p_{C} / 2$ each. 
It follows that the 
PMF 
of $W$ has a trinomial form \cite{Xia2020}, as depicted in Fig. \ref{fig:density_and_phase_psi_with_marked_vortices_and_KZ_scaling} (b):
\begin{equation}\label{P(W)_trinomial}
    P(W=n)=\sum_{\substack{m = |n| \\ m + n \, \text{even}}}^{\mathcal{N}}  \binom{\mathcal{N}}{m}\binom{m}{\frac{m+n}{2}}\left(\frac{\lambda}{2}\right)^{m}(1-\lambda)^{\mathcal{N}-m},
\end{equation}
with $\lambda=p_{v}p_{C}$.
The corresponding moment generating function is given by
\begin{equation}\label{moment_generating_function_W}
    \langle e^{Wt} \rangle=\prod_{i=1}^{\mathcal{N}} \langle e^{z_{i}t} \rangle=\left[1+\lambda\left(\cosh(t)-1\right)\right]^{\mathcal{N}},
\end{equation}
from which the moments of $W$ can be obtained as we detail in \cite{SM}. 
In particular, due to the symmetry of the distribution (\ref{P(W)_trinomial}), all odd moments vanish. Hence, we focus on the moments of $|W|$.
In the small-loop limit, when the maximum chord of $C$ is smaller than $\hat{\xi}$, every segment of length $\hat{\xi}$ with one endpoint inside the loop must cross its contour, giving $p_C = 2|A|/A_{\rm tot}$, with $|A|$ the enclosed area. Conversely, in the large loop limit, and provided that its curvature is sufficiently small, $p_{C}=2\hat{\xi}|C|/(\pi A_{tot})$. 
This yields the following scaling relations (see \cite{SM} for the full derivation):
\begin{equation}
\langle |W|^{p} \rangle \propto 
\begin{cases}
|A|/\hat{\xi}^{2}, & \text{for } |A| \ll \hat{\xi}^{2}, \\
(|C|/\hat{\xi})^{p/2}, & \text{for } |C| \gg \hat{\xi}, 
\end{cases}
\label{eq:W_every_powers_A}
\end{equation}
as verified by our numerical simulations for a circular loop as shown in Fig. \ref{fig:density_and_phase_psi_with_marked_vortices_and_KZ_scaling} (c).


\begin{figure}[t]
\includegraphics[width=\columnwidth]{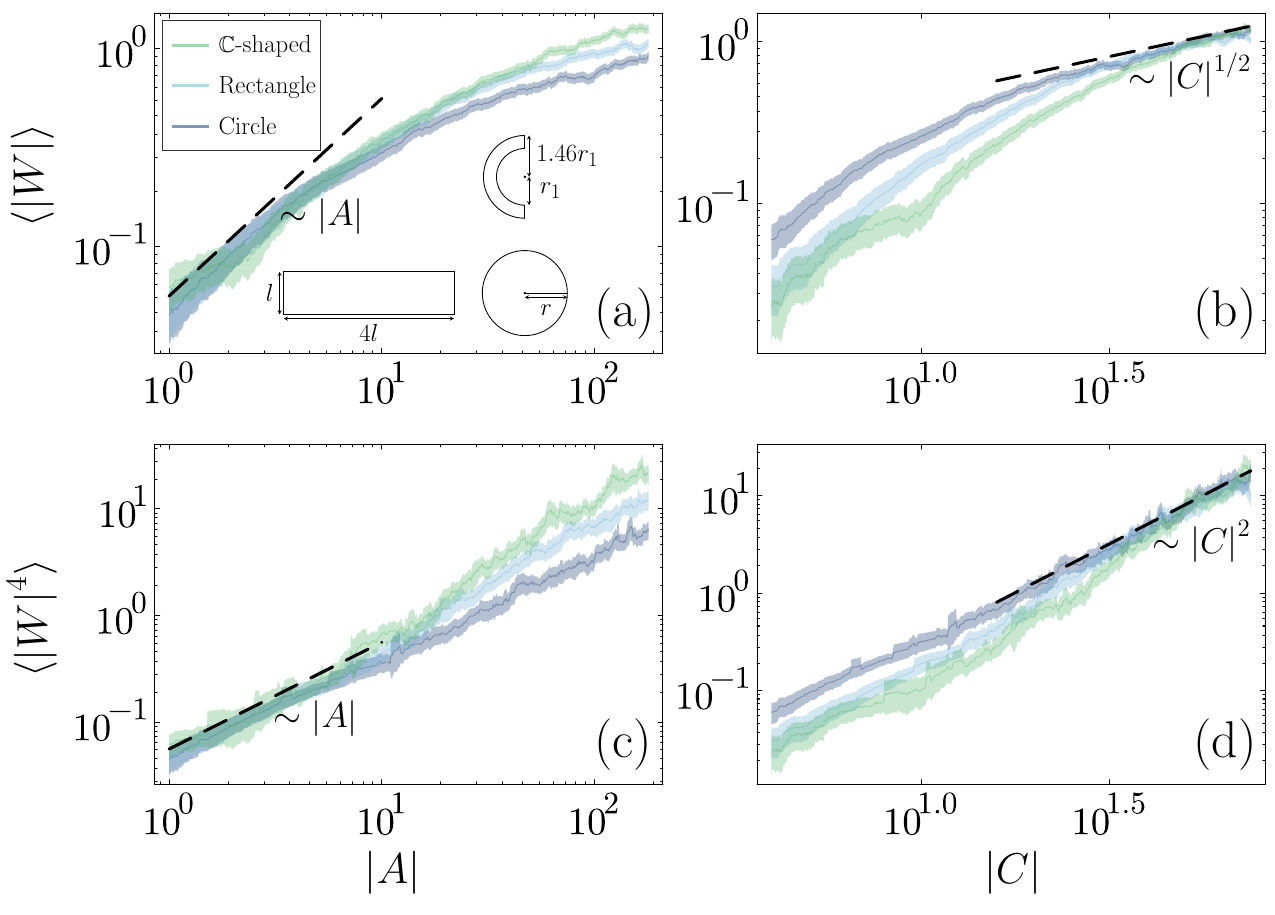}
\caption{\textbf{Moments of the absolute net vortex charge \(|W|\) for loops of various shapes.}  
The data correspond to a quench with \(\tau_Q=20\), averaged over \(\mathcal{R}=1000\) realizations; shaded bands indicate 95\% confidence intervals.  
Panels (a,b) show the first moment \(\langle|W|\rangle\) and panels (c,d) the fourth moment \(\langle|W|^2\rangle\), plotted respectively against enclosed area \(|A|\) in (a,c) and loop circumference \(|C|\) in (b,d).}
\label{fig:cumulants_Tq_10_fixed_area}
\end{figure}

\bigskip
{\it Testing the Migdal area rule.} 
The essence of the area rule is that, for loops within the inertial range (IR), the tails of the circulation 
probability density function (PDF) 
depend exclusively on the loop's enclosed area, not its shape. 
The first notable numerical test of this prediction in the context of classical turbulence is due to Iyer \emph{et al.} \cite{Iyer19}, who considered rectangular loops of different aspect ratios and observed that, for rectangles fully contained in the inertial range, the tails of the circulation PDF collapse regardless of the shape. Remarkably, this collapse seems to extend not only to the tails but also to the bulk of the PDF.
In the quantum scenario considered here, the inertial range spans from the typical vortex-core size up to the inter-vortex spacing \cite{Bradley12}. In the vortex pair model, the loop shape enters the circulation statistics exclusively through the crossing probability \(p_C\). As noted above, whenever the loop’s linear size remains below $\hat{\xi}$, i.e., within the IR, \(p_C\) depends solely on the enclosed area.
 
To test this prediction, we numerically computed the moments of $|W|$ for circular, rectangular, and $\mathbb{C}$-shaped loops $C$ (Fig.~\ref{fig:cumulants_Tq_10_fixed_area}). As anticipated, when plotted against loop area \(|A|\), all moments collapse onto a single line in the small-\(|A|\) (IR) limit, confirming shape independence. Additionally, in the large-\(|C|\) limit, the same moments, plotted versus the contour length \(|C|\), again fall onto a single line. This suggests that the circulation statistics does not depend on the loop shape for small loops, contained in the IR. To further test this statement, we fix the loop area and plot the full charge‐imbalance 
PMF 
for different loop geometries.
The result is shown in Fig.~\ref{fig:P(W)_different_shapes_fixed_area_collapse}. For loops outside the IR, the distributions for different shapes at fixed area do not collapse. In contrast, if the loop area is reduced so that all loops lie inside the IR, the distributions collapse, supporting the Migdal area rule.

\begin{figure}[ptb]
\includegraphics[width=0.82\columnwidth]{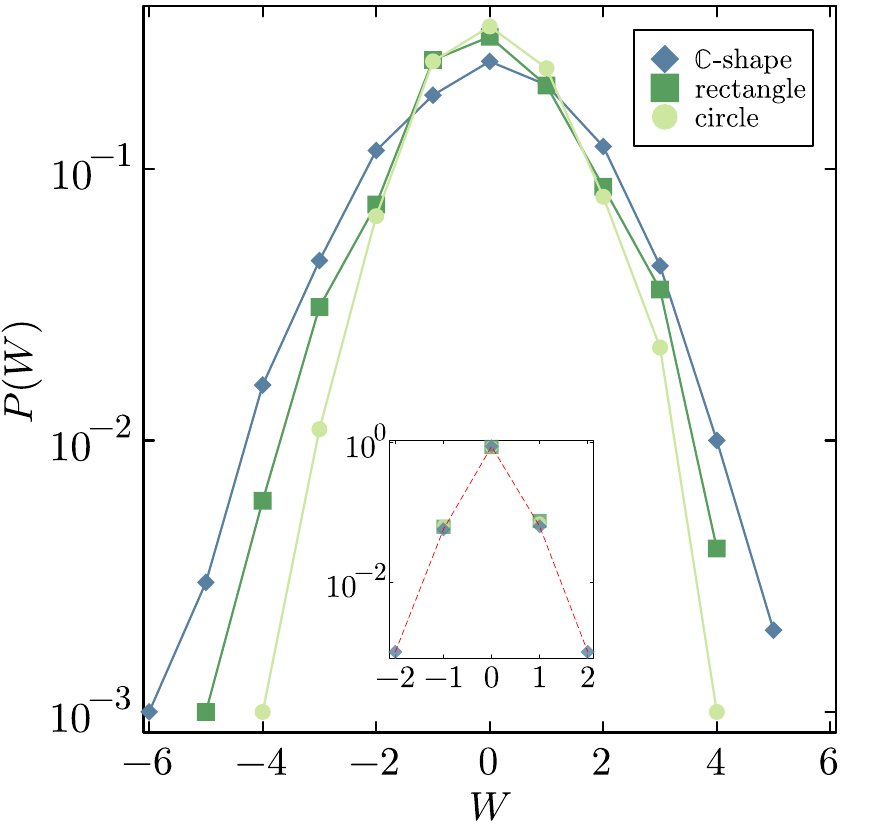}
\caption{\textbf{Net charge-imbalance distribution for loops of fixed area across different shapes.} The main plot shows the net vortex-charge 
PMF 
$P(W)$ at equilibration time after a quench of duration $\tau_Q=20$ for $\mathbb{C}$-shaped, rectangular, and circular loops of equal area $A=170$ outside the IR (see Fig.~\ref{fig:cumulants_Tq_10_fixed_area}). The inset shows the same distributions for loops of area $A=3$ inside the IR.
}
\label{fig:P(W)_different_shapes_fixed_area_collapse}
\end{figure}

{\it Kibble-Zurek universality of the circulation statistics.}
Having established the validity of the area rule, we now assess the Kibble–Zurek universality of the circulation statistics by fixing the shape of the loop and varying the quench rate. Combining $\hat\xi\propto\tau_Q^{1/4}$ with Eqs.~\eqref{eq:W_every_powers_A} gives $\langle|W|^p\rangle\propto |A|/\sqrt{\tau_Q}$ for small loops and $ \propto (|A|/\sqrt{\tau_Q})^{p/4}$ for large loops.
This is verified in Fig. (\ref{fig:cumulants_Tq_KZM_fixed_area}) for a circular contour, where the moments of $\langle |W| \rangle$ are plotted against the area $|A|$ for several $\tau_{Q}$ values. Rescaling the horizontal axis with $\hat{\xi}^{2}\propto\sqrt{\tau_{Q}}$ leads to a collapse of the moments onto a single curve, confirming the universal behavior of the circulation. 
Additionally, we report the moments of $|W|$ at fixed $|A|$ plotted against $\tau_Q$ in \cite{SM}. For large $|A|$, the newborn BEC complies with Zurek's prediction that the first moment scales as $\tau_Q^{-1/8}$  \cite{Zurek96c,Zurek13}. A fit to the numerical yields $\langle|W|\rangle\propto 
\tau_Q^{-0.147 \pm 0.018}$.
In contrast, in the small‐area limit, we observe 
$\langle|W|^p\rangle\propto
\tau_Q^{-0.502 \pm 0.040}
$ up to $p = 4$, supporting the KZM non-equilibrium universality of the area rule.

\begin{figure}[t]
\includegraphics[width=\columnwidth]{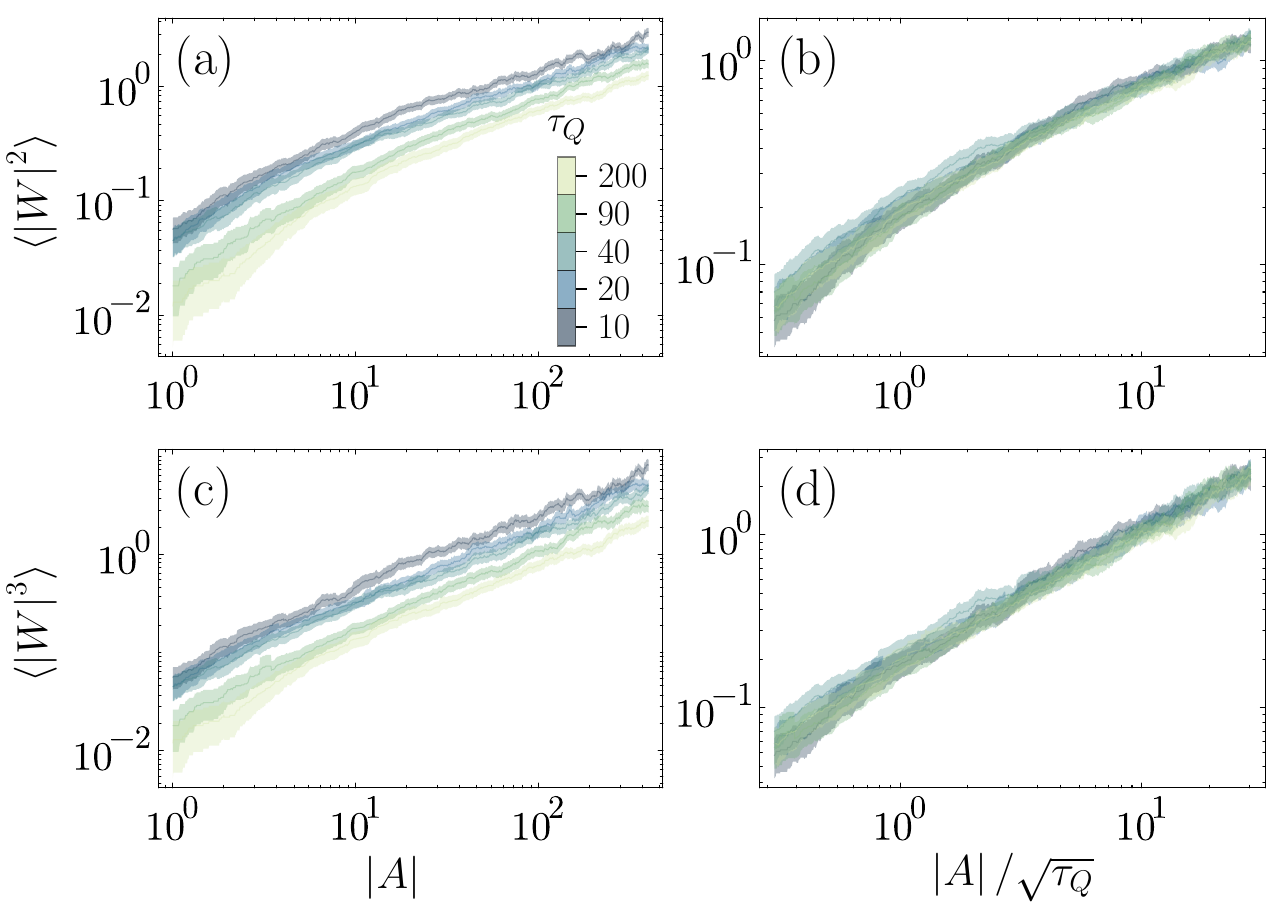}
\caption{\textbf{Kibble–Zurek universality of the net vortex charge $|W|$ in a circular contour of area $A$.} 
(a)–(b) show the second moment $\langle|W|^2\rangle$ plotted versus $|A|$ and $|A|/\sqrt{\tau_Q}$, respectively, for different values of the quench times $\tau_Q$ (averaged over $\mathcal{R}=1000$ realizations). 
(c)–(d) show the same for the third moment $\langle|W|^3\rangle$. 
Shaded error bands denote 95\% confidence intervals.}
\label{fig:cumulants_Tq_KZM_fixed_area}
\end{figure}

{\it Conclusion and outlooks.}
We have characterized the circulation statistics in a newborn Bose-Einstein condensate prepared by a thermal quench in finite time. Generalizing the Kibble-Zurek mechanism to account for the spatial distribution of spontaneously formed vortex–antivortex pairs, we have provided theoretical predictions for the circulation statistics within an area enclosed by a loop. Our results show that Migdal’s area rule—predicting the universality of circulation as a function of the enclosed area—holds for loops small compared to the typical vortex–antivortex pair separation, set by the universal KZM correlation length. We further uncover the universal scaling of the moments of the circulation statistics as a function of the quench time. 
Our work establishes a connection between the universal dynamics of critical phenomena and the topological underpinnings of spontaneous quantum turbulence. It offers testable predictions in systems undergoing $U(1)$ symmetry breaking, enabling experimental probes of vortex spatial statistics across platforms ranging from ultracold gases to multiferroics.

{\it Acknowledgments.}
It is a pleasure to thank Alexander Migdal, Fumika Suzuki, and Wojciech H. Zurek for insightful comments and discussions.
The authors acknowledge financial support from the Luxembourg National Research Fund under Grant No. C22/MS/17132060/BeyondKZM. MT would like to thank Anusandhan National Research Foundation (ANRF), Government of India, for the financial support through the Prime Minister Early Career Research Grant with Grant No. ANRF/ECRG/2024/003150/PMS.


\bibliographystyle{apsrev4}
\let\itshape\upshape
\normalem
\bibliography{defects_QT}

\providecommand{\noopsort}[1]{}\providecommand{\singleletter}[1]{#1}%
\begin{thebibliography}{52}%
\makeatletter
\providecommand \@ifxundefined [1]{%
 \@ifx{#1\undefined}
}%
\providecommand \@ifnum [1]{%
 \ifnum #1\expandafter \@firstoftwo
 \else \expandafter \@secondoftwo
 \fi
}%
\providecommand \@ifx [1]{%
 \ifx #1\expandafter \@firstoftwo
 \else \expandafter \@secondoftwo
 \fi
}%
\providecommand \natexlab [1]{#1}%
\providecommand \enquote  [1]{``#1''}%
\providecommand \bibnamefont  [1]{#1}%
\providecommand \bibfnamefont [1]{#1}%
\providecommand \citenamefont [1]{#1}%
\providecommand \href@noop [0]{\@secondoftwo}%
\providecommand \href [0]{\begingroup \@sanitize@url \@href}%
\providecommand \@href[1]{\@@startlink{#1}\@@href}%
\providecommand \@@href[1]{\endgroup#1\@@endlink}%
\providecommand \@sanitize@url [0]{\catcode `\\12\catcode `\$12\catcode `\&12\catcode `\#12\catcode `\^12\catcode `\_12\catcode `\%12\relax}%
\providecommand \@@startlink[1]{}%
\providecommand \@@endlink[0]{}%
\providecommand \url  [0]{\begingroup\@sanitize@url \@url }%
\providecommand \@url [1]{\endgroup\@href {#1}{\urlprefix }}%
\providecommand \urlprefix  [0]{URL }%
\providecommand \Eprint [0]{\href }%
\providecommand \doibase [0]{http://dx.doi.org/}%
\providecommand \selectlanguage [0]{\@gobble}%
\providecommand \bibinfo  [0]{\@secondoftwo}%
\providecommand \bibfield  [0]{\@secondoftwo}%
\providecommand \translation [1]{[#1]}%
\providecommand \BibitemOpen [0]{}%
\providecommand \bibitemStop [0]{}%
\providecommand \bibitemNoStop [0]{.\EOS\space}%
\providecommand \EOS [0]{\spacefactor3000\relax}%
\providecommand \BibitemShut  [1]{\csname bibitem#1\endcsname}%
\let\auto@bib@innerbib\@empty
\bibitem [{\citenamefont {Migdal}(1994)}]{Migdal94}%
  \BibitemOpen
  \bibfield  {author} {\bibinfo {author} {\bibfnamefont {A.~A.}\ \bibnamefont {Migdal}},\ }\bibfield  {title} {\enquote {\bibinfo {title} {Loop equation and area law in turbulence},}\ }\href {\doibase 10.1142/S0217751X94000558} {\bibfield  {journal} {\bibinfo  {journal} {International Journal of Modern Physics A}\ }\textbf {\bibinfo {volume} {09}},\ \bibinfo {pages} {1197} (\bibinfo {year} {1994})}\BibitemShut {NoStop}%
\bibitem [{\citenamefont {Migdal}(2019{\natexlab{a}})}]{migdal2019universal}%
  \BibitemOpen
  \bibfield  {author} {\bibinfo {author} {\bibfnamefont {A.}~\bibnamefont {Migdal}},\ }\bibfield  {title} {\enquote {\bibinfo {title} {Universal area law in turbulence},}\ }\href {https://arxiv.org/abs/1903.08613} {\  (\bibinfo {year} {2019}{\natexlab{a}})},\ \Eprint {http://arxiv.org/abs/1903.08613} {arXiv:1903.08613 [hep-th]} \BibitemShut {NoStop}%
\bibitem [{\citenamefont {Migdal}(2019{\natexlab{b}})}]{migdal2019exact}%
  \BibitemOpen
  \bibfield  {author} {\bibinfo {author} {\bibfnamefont {A.}~\bibnamefont {Migdal}},\ }\bibfield  {title} {\enquote {\bibinfo {title} {Exact area law for planar loops in turbulence in two and three dimensions},}\ }\href {https://arxiv.org/abs/1904.05245} {\  (\bibinfo {year} {2019}{\natexlab{b}})},\ \Eprint {http://arxiv.org/abs/1904.05245} {arXiv:1904.05245 [hep-th]} \BibitemShut {NoStop}%
\bibitem [{\citenamefont {Iyer}\ \emph {et~al.}(2019)\citenamefont {Iyer}, \citenamefont {Sreenivasan},\ and\ \citenamefont {Yeung}}]{Iyer19}%
  \BibitemOpen
  \bibfield  {author} {\bibinfo {author} {\bibfnamefont {K.~P.}\ \bibnamefont {Iyer}}, \bibinfo {author} {\bibfnamefont {K.~R.}\ \bibnamefont {Sreenivasan}}, \ and\ \bibinfo {author} {\bibfnamefont {P.~K.}\ \bibnamefont {Yeung}},\ }\bibfield  {title} {\enquote {\bibinfo {title} {Circulation in high reynolds number isotropic turbulence is a bifractal},}\ }\href {\doibase 10.1103/PhysRevX.9.041006} {\bibfield  {journal} {\bibinfo  {journal} {Phys. Rev. X}\ }\textbf {\bibinfo {volume} {9}},\ \bibinfo {pages} {041006} (\bibinfo {year} {2019})}\BibitemShut {NoStop}%
\bibitem [{\citenamefont {Apolin\'ario}\ \emph {et~al.}(2020)\citenamefont {Apolin\'ario}, \citenamefont {Moriconi}, \citenamefont {Pereira},\ and\ \citenamefont {Valad\~ao}}]{Apolinario20}%
  \BibitemOpen
  \bibfield  {author} {\bibinfo {author} {\bibfnamefont {G.~B.}\ \bibnamefont {Apolin\'ario}}, \bibinfo {author} {\bibfnamefont {L.}~\bibnamefont {Moriconi}}, \bibinfo {author} {\bibfnamefont {R.~M.}\ \bibnamefont {Pereira}}, \ and\ \bibinfo {author} {\bibfnamefont {V.~J.}\ \bibnamefont {Valad\~ao}},\ }\bibfield  {title} {\enquote {\bibinfo {title} {Vortex gas modeling of turbulent circulation statistics},}\ }\href {\doibase 10.1103/PhysRevE.102.041102} {\bibfield  {journal} {\bibinfo  {journal} {Phys. Rev. E}\ }\textbf {\bibinfo {volume} {102}},\ \bibinfo {pages} {041102} (\bibinfo {year} {2020})}\BibitemShut {NoStop}%
\bibitem [{\citenamefont {Iyer}\ \emph {et~al.}(2021)\citenamefont {Iyer}, \citenamefont {Bharadwaj},\ and\ \citenamefont {Sreenivasan}}]{Iyer21}%
  \BibitemOpen
  \bibfield  {author} {\bibinfo {author} {\bibfnamefont {K.~P.}\ \bibnamefont {Iyer}}, \bibinfo {author} {\bibfnamefont {S.~S.}\ \bibnamefont {Bharadwaj}}, \ and\ \bibinfo {author} {\bibfnamefont {K.~R.}\ \bibnamefont {Sreenivasan}},\ }\bibfield  {title} {\enquote {\bibinfo {title} {The area rule for circulation in three-dimensional turbulence},}\ }\href {\doibase 10.1073/pnas.2114679118} {\bibfield  {journal} {\bibinfo  {journal} {Proceedings of the National Academy of Sciences}\ }\textbf {\bibinfo {volume} {118}},\ \bibinfo {pages} {e2114679118} (\bibinfo {year} {2021})}\BibitemShut {NoStop}%
\bibitem [{\citenamefont {M\"uller}\ \emph {et~al.}(2021)\citenamefont {M\"uller}, \citenamefont {Polanco},\ and\ \citenamefont {Krstulovic}}]{Muller21}%
  \BibitemOpen
  \bibfield  {author} {\bibinfo {author} {\bibfnamefont {N.~P.}\ \bibnamefont {M\"uller}}, \bibinfo {author} {\bibfnamefont {J.~I.}\ \bibnamefont {Polanco}}, \ and\ \bibinfo {author} {\bibfnamefont {G.}~\bibnamefont {Krstulovic}},\ }\bibfield  {title} {\enquote {\bibinfo {title} {Intermittency of velocity circulation in quantum turbulence},}\ }\href {\doibase 10.1103/PhysRevX.11.011053} {\bibfield  {journal} {\bibinfo  {journal} {Phys. Rev. X}\ }\textbf {\bibinfo {volume} {11}},\ \bibinfo {pages} {011053} (\bibinfo {year} {2021})}\BibitemShut {NoStop}%
\bibitem [{\citenamefont {Xie}\ and\ \citenamefont {Xie}(2025)}]{Xie2025}%
  \BibitemOpen
  \bibfield  {author} {\bibinfo {author} {\bibfnamefont {B.-J.}\ \bibnamefont {Xie}}\ and\ \bibinfo {author} {\bibfnamefont {J.-H.}\ \bibnamefont {Xie}},\ }\href {https://arxiv.org/abs/2504.21512} {\enquote {\bibinfo {title} {Area rule of velocity circulation in two-dimensional instability-driven turbulence beyond the inertial range},}\ } (\bibinfo {year} {2025}),\ \Eprint {http://arxiv.org/abs/2504.21512} {arXiv:2504.21512 [physics.flu-dyn]} \BibitemShut {NoStop}%
\bibitem [{\citenamefont {Zhu}\ \emph {et~al.}(2023)\citenamefont {Zhu}, \citenamefont {Xie},\ and\ \citenamefont {Xia}}]{Zhu23}%
  \BibitemOpen
  \bibfield  {author} {\bibinfo {author} {\bibfnamefont {H.-Y.}\ \bibnamefont {Zhu}}, \bibinfo {author} {\bibfnamefont {J.-H.}\ \bibnamefont {Xie}}, \ and\ \bibinfo {author} {\bibfnamefont {K.-Q.}\ \bibnamefont {Xia}},\ }\bibfield  {title} {\enquote {\bibinfo {title} {Circulation in quasi-2d turbulence: Experimental observation of the area rule and bifractality},}\ }\href {\doibase 10.1103/PhysRevLett.130.214001} {\bibfield  {journal} {\bibinfo  {journal} {Phys. Rev. Lett.}\ }\textbf {\bibinfo {volume} {130}},\ \bibinfo {pages} {214001} (\bibinfo {year} {2023})}\BibitemShut {NoStop}%
\bibitem [{\citenamefont {Tsatsos}\ \emph {et~al.}(2016)\citenamefont {Tsatsos}, \citenamefont {Tavares}, \citenamefont {Cidrim}, \citenamefont {Fritsch}, \citenamefont {Caracanhas}, \citenamefont {{dos Santos}}, \citenamefont {Barenghi},\ and\ \citenamefont {Bagnato}}]{Tsatsos16}%
  \BibitemOpen
  \bibfield  {author} {\bibinfo {author} {\bibfnamefont {M.~C.}\ \bibnamefont {Tsatsos}}, \bibinfo {author} {\bibfnamefont {P.~E.}\ \bibnamefont {Tavares}}, \bibinfo {author} {\bibfnamefont {A.}~\bibnamefont {Cidrim}}, \bibinfo {author} {\bibfnamefont {A.~R.}\ \bibnamefont {Fritsch}}, \bibinfo {author} {\bibfnamefont {M.~A.}\ \bibnamefont {Caracanhas}}, \bibinfo {author} {\bibfnamefont {F.~E.~A.}\ \bibnamefont {{dos Santos}}}, \bibinfo {author} {\bibfnamefont {C.~F.}\ \bibnamefont {Barenghi}}, \ and\ \bibinfo {author} {\bibfnamefont {V.~S.}\ \bibnamefont {Bagnato}},\ }\bibfield  {title} {\enquote {\bibinfo {title} {Quantum turbulence in trapped atomic bose–einstein condensates},}\ }\href {\doibase https://doi.org/10.1016/j.physrep.2016.02.003} {\bibfield  {journal} {\bibinfo  {journal} {Physics Reports}\ }\textbf {\bibinfo {volume} {622}},\ \bibinfo {pages} {1} (\bibinfo {year} {2016})}\BibitemShut {NoStop}%
\bibitem [{\citenamefont {Polanco}\ \emph {et~al.}(2021)\citenamefont {Polanco}, \citenamefont {M{\"u}ller},\ and\ \citenamefont {Krstulovic}}]{Polanco21}%
  \BibitemOpen
  \bibfield  {author} {\bibinfo {author} {\bibfnamefont {J.~I.}\ \bibnamefont {Polanco}}, \bibinfo {author} {\bibfnamefont {N.~P.}\ \bibnamefont {M{\"u}ller}}, \ and\ \bibinfo {author} {\bibfnamefont {G.}~\bibnamefont {Krstulovic}},\ }\bibfield  {title} {\enquote {\bibinfo {title} {Vortex clustering, polarisation and circulation intermittency in classical and quantum turbulence},}\ }\href {\doibase 10.1038/s41467-021-27382-6} {\bibfield  {journal} {\bibinfo  {journal} {Nature Communications}\ }\textbf {\bibinfo {volume} {12}},\ \bibinfo {pages} {7090} (\bibinfo {year} {2021})}\BibitemShut {NoStop}%
\bibitem [{\citenamefont {Weiler}\ \emph {et~al.}(2008)\citenamefont {Weiler}, \citenamefont {Neely}, \citenamefont {Scherer}, \citenamefont {Bradley}, \citenamefont {Davis},\ and\ \citenamefont {Anderson}}]{Weiler08}%
  \BibitemOpen
  \bibfield  {author} {\bibinfo {author} {\bibfnamefont {C.~N.}\ \bibnamefont {Weiler}}, \bibinfo {author} {\bibfnamefont {T.~W.}\ \bibnamefont {Neely}}, \bibinfo {author} {\bibfnamefont {D.~R.}\ \bibnamefont {Scherer}}, \bibinfo {author} {\bibfnamefont {A.~S.}\ \bibnamefont {Bradley}}, \bibinfo {author} {\bibfnamefont {M.~J.}\ \bibnamefont {Davis}}, \ and\ \bibinfo {author} {\bibfnamefont {B.~P.}\ \bibnamefont {Anderson}},\ }\bibfield  {title} {\enquote {\bibinfo {title} {Spontaneous vortices in the formation of {B}ose-{E}instein condensates},}\ }\href {\doibase 10.1038/nature07334} {\bibfield  {journal} {\bibinfo  {journal} {Nature}\ }\textbf {\bibinfo {volume} {455}},\ \bibinfo {pages} {948} (\bibinfo {year} {2008})}\BibitemShut {NoStop}%
\bibitem [{\citenamefont {del Campo}\ \emph {et~al.}(2011)\citenamefont {del Campo}, \citenamefont {Retzker},\ and\ \citenamefont {Plenio}}]{DRP11}%
  \BibitemOpen
  \bibfield  {author} {\bibinfo {author} {\bibfnamefont {A.}~\bibnamefont {del Campo}}, \bibinfo {author} {\bibfnamefont {A.}~\bibnamefont {Retzker}}, \ and\ \bibinfo {author} {\bibfnamefont {M.~B.}\ \bibnamefont {Plenio}},\ }\bibfield  {title} {\enquote {\bibinfo {title} {The inhomogeneous {K}ibble-{Z}urek mechanism: vortex nucleation during {B}ose-{E}instein condensation},}\ }\href {http://stacks.iop.org/1367-2630/13/i=8/a=083022} {\bibfield  {journal} {\bibinfo  {journal} {\href{http://stacks.iop.org/1367-2630/13/i=8/a=083022}{New J. Phys.}}\ }\textbf {\bibinfo {volume} {13}},\ \bibinfo {pages} {083022} (\bibinfo {year} {2011})}\BibitemShut {NoStop}%
\bibitem [{\citenamefont {Chesler}\ \emph {et~al.}(2015)\citenamefont {Chesler}, \citenamefont {Garc\'{\i}a-Garc\'{\i}a},\ and\ \citenamefont {Liu}}]{Chesler15}%
  \BibitemOpen
  \bibfield  {author} {\bibinfo {author} {\bibfnamefont {P.~M.}\ \bibnamefont {Chesler}}, \bibinfo {author} {\bibfnamefont {A.~M.}\ \bibnamefont {Garc\'{\i}a-Garc\'{\i}a}}, \ and\ \bibinfo {author} {\bibfnamefont {H.}~\bibnamefont {Liu}},\ }\bibfield  {title} {\enquote {\bibinfo {title} {Defect formation beyond kibble-zurek mechanism and holography},}\ }\href {\doibase 10.1103/PhysRevX.5.021015} {\bibfield  {journal} {\bibinfo  {journal} {Phys. Rev. X}\ }\textbf {\bibinfo {volume} {5}},\ \bibinfo {pages} {021015} (\bibinfo {year} {2015})}\BibitemShut {NoStop}%
\bibitem [{\citenamefont {Zeng}\ \emph {et~al.}(2023)\citenamefont {Zeng}, \citenamefont {Xia},\ and\ \citenamefont {del Campo}}]{Zeng23}%
  \BibitemOpen
  \bibfield  {author} {\bibinfo {author} {\bibfnamefont {H.-B.}\ \bibnamefont {Zeng}}, \bibinfo {author} {\bibfnamefont {C.-Y.}\ \bibnamefont {Xia}}, \ and\ \bibinfo {author} {\bibfnamefont {A.}~\bibnamefont {del Campo}},\ }\bibfield  {title} {\enquote {\bibinfo {title} {Universal breakdown of kibble-zurek scaling in fast quenches across a phase transition},}\ }\href {\doibase 10.1103/PhysRevLett.130.060402} {\bibfield  {journal} {\bibinfo  {journal} {Phys. Rev. Lett.}\ }\textbf {\bibinfo {volume} {130}},\ \bibinfo {pages} {060402} (\bibinfo {year} {2023})}\BibitemShut {NoStop}%
\bibitem [{\citenamefont {Thudiyangal}\ and\ \citenamefont {del Campo}(2024)}]{Thudiyangal24}%
  \BibitemOpen
  \bibfield  {author} {\bibinfo {author} {\bibfnamefont {M.}~\bibnamefont {Thudiyangal}}\ and\ \bibinfo {author} {\bibfnamefont {A.}~\bibnamefont {del Campo}},\ }\bibfield  {title} {\enquote {\bibinfo {title} {Universal vortex statistics and stochastic geometry of bose-einstein condensation},}\ }\href {\doibase 10.1103/PhysRevResearch.6.033152} {\bibfield  {journal} {\bibinfo  {journal} {Phys. Rev. Res.}\ }\textbf {\bibinfo {volume} {6}},\ \bibinfo {pages} {033152} (\bibinfo {year} {2024})}\BibitemShut {NoStop}%
\bibitem [{\citenamefont {Kibble}(1976)}]{Kibble76a}%
  \BibitemOpen
  \bibfield  {author} {\bibinfo {author} {\bibfnamefont {T.~W.~B.}\ \bibnamefont {Kibble}},\ }\bibfield  {title} {\enquote {\bibinfo {title} {Topology of cosmic domains and strings},}\ }\href {http://stacks.iop.org/0305-4470/9/i=8/a=029} {\bibfield  {journal} {\bibinfo  {journal} {\href{http://stacks.iop.org/0305-4470/9/i=8/a=029}{J. of Phys. A: Math. Gen.}}\ }\textbf {\bibinfo {volume} {9}},\ \bibinfo {pages} {1387} (\bibinfo {year} {1976})}\BibitemShut {NoStop}%
\bibitem [{\citenamefont {Kibble}(1980)}]{Kibble76b}%
  \BibitemOpen
  \bibfield  {author} {\bibinfo {author} {\bibfnamefont {T.~W.~B.}\ \bibnamefont {Kibble}},\ }\bibfield  {title} {\enquote {\bibinfo {title} {Some implications of a cosmological phase transition},}\ }\href {http://www.sciencedirect.com/science/article/pii/0370157380900915} {\bibfield  {journal} {\bibinfo  {journal} {\href{http://www.sciencedirect.com/science/article/pii/0370157380900915}{Phys. Reports}}\ }\textbf {\bibinfo {volume} {67}},\ \bibinfo {pages} {183} (\bibinfo {year} {1980})}\BibitemShut {NoStop}%
\bibitem [{\citenamefont {Zurek}(1985)}]{Zurek96a}%
  \BibitemOpen
  \bibfield  {author} {\bibinfo {author} {\bibfnamefont {W.~H.}\ \bibnamefont {Zurek}},\ }\bibfield  {title} {\enquote {\bibinfo {title} {Cosmological experiments in superfluid helium?}}\ }\href {http://dx.doi.org/10.1038/317505a0} {\bibfield  {journal} {\bibinfo  {journal} {\href{http://dx.doi.org/10.1038/317505a0}{Nature}}\ }\textbf {\bibinfo {volume} {317}},\ \bibinfo {pages} {505} (\bibinfo {year} {1985})}\BibitemShut {NoStop}%
\bibitem [{\citenamefont {Zurek}(1993{\natexlab{a}})}]{Zurek96b}%
  \BibitemOpen
  \bibfield  {author} {\bibinfo {author} {\bibfnamefont {W.~H.}\ \bibnamefont {Zurek}},\ }\bibfield  {title} {\enquote {\bibinfo {title} {Cosmic strings in laboratory superfluids and the topological remnants of other phase transitions},}\ }\href {http://www.actaphys.uj.edu.pl/fulltext?series=Reg&vol=24&page=1301} {\bibfield  {journal} {\bibinfo  {journal} {\href{http://www.actaphys.uj.edu.pl/fulltext?series=Reg&vol=24&page=1301}{Acta Phys. Pol. B}}\ }\textbf {\bibinfo {volume} {24}},\ \bibinfo {pages} {1301} (\bibinfo {year} {1993}{\natexlab{a}})}\BibitemShut {NoStop}%
\bibitem [{\citenamefont {del Campo}\ and\ \citenamefont {Zurek}(2014)}]{DZ14}%
  \BibitemOpen
  \bibfield  {author} {\bibinfo {author} {\bibfnamefont {A.}~\bibnamefont {del Campo}}\ and\ \bibinfo {author} {\bibfnamefont {W.~H.}\ \bibnamefont {Zurek}},\ }\bibfield  {title} {\enquote {\bibinfo {title} {Universality of phase transition dynamics: Topological defects from symmetry breaking},}\ }\href {\doibase 10.1142/S0217751X1430018X} {\bibfield  {journal} {\bibinfo  {journal} {Int. J. Mod. Phys. A}\ }\textbf {\bibinfo {volume} {29}},\ \bibinfo {pages} {1430018} (\bibinfo {year} {2014})}\BibitemShut {NoStop}%
\bibitem [{\citenamefont {can Yang}(2025)}]{Yang2025}%
  \BibitemOpen
  \bibfield  {author} {\bibinfo {author} {\bibfnamefont {W.}~\bibnamefont {can Yang}},\ }\href {https://arxiv.org/abs/2504.04409} {\enquote {\bibinfo {title} {Non-equilibrium dynamics and universality of 4d quantum vortices and turbulence},}\ } (\bibinfo {year} {2025}),\ \Eprint {http://arxiv.org/abs/2504.04409} {arXiv:2504.04409 [cond-mat.quant-gas]} \BibitemShut {NoStop}%
\bibitem [{\citenamefont {Shinn}\ \emph {et~al.}(2025)\citenamefont {Shinn}, \citenamefont {Massaro}, \citenamefont {Thudiyangal},\ and\ \citenamefont {del Campo}}]{shinn2025}%
  \BibitemOpen
  \bibfield  {author} {\bibinfo {author} {\bibfnamefont {S.-H.}\ \bibnamefont {Shinn}}, \bibinfo {author} {\bibfnamefont {M.}~\bibnamefont {Massaro}}, \bibinfo {author} {\bibfnamefont {M.}~\bibnamefont {Thudiyangal}}, \ and\ \bibinfo {author} {\bibfnamefont {A.}~\bibnamefont {del Campo}},\ }\href {https://arxiv.org/abs/2506.21670} {\enquote {\bibinfo {title} {Spontaneous quantum turbulence in a newborn bose-einstein condensate via the kibble-zurek mechanism},}\ } (\bibinfo {year} {2025}),\ \Eprint {http://arxiv.org/abs/2506.21670} {arXiv:2506.21670 [cond-mat.quant-gas]} \BibitemShut {NoStop}%
\bibitem [{\citenamefont {Tsubota}\ \emph {et~al.}(2013)\citenamefont {Tsubota}, \citenamefont {Kobayashi},\ and\ \citenamefont {Takeuchi}}]{Tsubota13}%
  \BibitemOpen
  \bibfield  {author} {\bibinfo {author} {\bibfnamefont {M.}~\bibnamefont {Tsubota}}, \bibinfo {author} {\bibfnamefont {M.}~\bibnamefont {Kobayashi}}, \ and\ \bibinfo {author} {\bibfnamefont {H.}~\bibnamefont {Takeuchi}},\ }\bibfield  {title} {\enquote {\bibinfo {title} {Quantum hydrodynamics},}\ }\href {\doibase https://doi.org/10.1016/j.physrep.2012.09.007} {\bibfield  {journal} {\bibinfo  {journal} {Physics Reports}\ }\textbf {\bibinfo {volume} {522}},\ \bibinfo {pages} {191} (\bibinfo {year} {2013})},\ \bibinfo {note} {quantum hydrodynamics}\BibitemShut {NoStop}%
\bibitem [{\citenamefont {Zurek}(1993{\natexlab{b}})}]{Zurek96c}%
  \BibitemOpen
  \bibfield  {author} {\bibinfo {author} {\bibfnamefont {W.~H.}\ \bibnamefont {Zurek}},\ }\bibfield  {title} {\enquote {\bibinfo {title} {Cosmological experiments in condensed matter systems},}\ }\href {http://www.sciencedirect.com/science/article/pii/S0370157396000099} {\bibfield  {journal} {\bibinfo  {journal} {\href{http://www.sciencedirect.com/science/article/pii/S0370157396000099}{Phys. Reports}}\ }\textbf {\bibinfo {volume} {276}},\ \bibinfo {pages} {177} (\bibinfo {year} {1993}{\natexlab{b}})}\BibitemShut {NoStop}%
\bibitem [{\citenamefont {Monaco}\ \emph {et~al.}(2002)\citenamefont {Monaco}, \citenamefont {Mygind},\ and\ \citenamefont {Rivers}}]{Monaco02}%
  \BibitemOpen
  \bibfield  {author} {\bibinfo {author} {\bibfnamefont {R.}~\bibnamefont {Monaco}}, \bibinfo {author} {\bibfnamefont {J.}~\bibnamefont {Mygind}}, \ and\ \bibinfo {author} {\bibfnamefont {R.~J.}\ \bibnamefont {Rivers}},\ }\bibfield  {title} {\enquote {\bibinfo {title} {{Z}urek-{K}ibble domain structures: The dynamics of spontaneous vortex formation in annular josephson tunnel junctions},}\ }\href {\doibase 10.1103/PhysRevLett.89.080603} {\bibfield  {journal} {\bibinfo  {journal} {Phys. Rev. Lett.}\ }\textbf {\bibinfo {volume} {89}},\ \bibinfo {pages} {080603} (\bibinfo {year} {2002})}\BibitemShut {NoStop}%
\bibitem [{\citenamefont {Monaco}\ \emph {et~al.}(2003)\citenamefont {Monaco}, \citenamefont {Mygind},\ and\ \citenamefont {Rivers}}]{Monaco03}%
  \BibitemOpen
  \bibfield  {author} {\bibinfo {author} {\bibfnamefont {R.}~\bibnamefont {Monaco}}, \bibinfo {author} {\bibfnamefont {J.}~\bibnamefont {Mygind}}, \ and\ \bibinfo {author} {\bibfnamefont {R.~J.}\ \bibnamefont {Rivers}},\ }\bibfield  {title} {\enquote {\bibinfo {title} {Spontaneous fluxon formation in annular josephson tunnel junctions},}\ }\href {\doibase 10.1103/PhysRevB.67.104506} {\bibfield  {journal} {\bibinfo  {journal} {Phys. Rev. B}\ }\textbf {\bibinfo {volume} {67}},\ \bibinfo {pages} {104506} (\bibinfo {year} {2003})}\BibitemShut {NoStop}%
\bibitem [{\citenamefont {Monaco}\ \emph {et~al.}(2006)\citenamefont {Monaco}, \citenamefont {Aaroe}, \citenamefont {Mygind}, \citenamefont {Rivers},\ and\ \citenamefont {Koshelets}}]{Monaco06}%
  \BibitemOpen
  \bibfield  {author} {\bibinfo {author} {\bibfnamefont {R.}~\bibnamefont {Monaco}}, \bibinfo {author} {\bibfnamefont {M.}~\bibnamefont {Aaroe}}, \bibinfo {author} {\bibfnamefont {J.}~\bibnamefont {Mygind}}, \bibinfo {author} {\bibfnamefont {R.~J.}\ \bibnamefont {Rivers}}, \ and\ \bibinfo {author} {\bibfnamefont {V.~P.}\ \bibnamefont {Koshelets}},\ }\bibfield  {title} {\enquote {\bibinfo {title} {Experiments on spontaneous vortex formation in josephson tunnel junctions},}\ }\href {\doibase 10.1103/PhysRevB.74.144513} {\bibfield  {journal} {\bibinfo  {journal} {Phys. Rev. B}\ }\textbf {\bibinfo {volume} {74}},\ \bibinfo {pages} {144513} (\bibinfo {year} {2006})}\BibitemShut {NoStop}%
\bibitem [{\citenamefont {Monaco}\ \emph {et~al.}(2009)\citenamefont {Monaco}, \citenamefont {Mygind}, \citenamefont {Rivers},\ and\ \citenamefont {Koshelets}}]{Monaco09}%
  \BibitemOpen
  \bibfield  {author} {\bibinfo {author} {\bibfnamefont {R.}~\bibnamefont {Monaco}}, \bibinfo {author} {\bibfnamefont {J.}~\bibnamefont {Mygind}}, \bibinfo {author} {\bibfnamefont {R.~J.}\ \bibnamefont {Rivers}}, \ and\ \bibinfo {author} {\bibfnamefont {V.~P.}\ \bibnamefont {Koshelets}},\ }\bibfield  {title} {\enquote {\bibinfo {title} {Spontaneous fluxoid formation in superconducting loops},}\ }\href {\doibase 10.1103/PhysRevB.80.180501} {\bibfield  {journal} {\bibinfo  {journal} {Phys. Rev. B}\ }\textbf {\bibinfo {volume} {80}},\ \bibinfo {pages} {180501} (\bibinfo {year} {2009})}\BibitemShut {NoStop}%
\bibitem [{\citenamefont {Das}\ \emph {et~al.}(2012)\citenamefont {Das}, \citenamefont {Sabbatini},\ and\ \citenamefont {Zurek}}]{Das12}%
  \BibitemOpen
  \bibfield  {author} {\bibinfo {author} {\bibfnamefont {A.}~\bibnamefont {Das}}, \bibinfo {author} {\bibfnamefont {J.}~\bibnamefont {Sabbatini}}, \ and\ \bibinfo {author} {\bibfnamefont {W.~H.}\ \bibnamefont {Zurek}},\ }\bibfield  {title} {\enquote {\bibinfo {title} {Winding up superfluid in a torus via {B}ose-{E}instein condensation},}\ }\href {\doibase 10.1038/srep00352} {\bibfield  {journal} {\bibinfo  {journal} {Sci. Rep.}\ }\textbf {\bibinfo {volume} {2}},\ \bibinfo {pages} {352} (\bibinfo {year} {2012})}\BibitemShut {NoStop}%
\bibitem [{\citenamefont {Zurek}(2013)}]{Zurek13}%
  \BibitemOpen
  \bibfield  {author} {\bibinfo {author} {\bibfnamefont {W.~H.}\ \bibnamefont {Zurek}},\ }\bibfield  {title} {\enquote {\bibinfo {title} {Topological relics of symmetry breaking: winding numbers and scaling tilts from random vortex{\textendash}antivortex pairs},}\ }\href {\doibase 10.1088/0953-8984/25/40/404209} {\bibfield  {journal} {\bibinfo  {journal} {Journal of Physics: Condensed Matter}\ }\textbf {\bibinfo {volume} {25}},\ \bibinfo {pages} {404209} (\bibinfo {year} {2013})}\BibitemShut {NoStop}%
\bibitem [{\citenamefont {Sonner}\ \emph {et~al.}(2015)\citenamefont {Sonner}, \citenamefont {del Campo},\ and\ \citenamefont {Zurek}}]{Sonner15}%
  \BibitemOpen
  \bibfield  {author} {\bibinfo {author} {\bibfnamefont {J.}~\bibnamefont {Sonner}}, \bibinfo {author} {\bibfnamefont {A.}~\bibnamefont {del Campo}}, \ and\ \bibinfo {author} {\bibfnamefont {W.~H.}\ \bibnamefont {Zurek}},\ }\bibfield  {title} {\enquote {\bibinfo {title} {Universal far-from-equilibrium dynamics of a holographic superconductor},}\ }\href {\doibase 10.1038/ncomms8406} {\bibfield  {journal} {\bibinfo  {journal} {Nature Communications}\ }\textbf {\bibinfo {volume} {6}},\ \bibinfo {pages} {7406} (\bibinfo {year} {2015})}\BibitemShut {NoStop}%
\bibitem [{\citenamefont {Zeng}\ \emph {et~al.}(2021)\citenamefont {Zeng}, \citenamefont {Xia},\ and\ \citenamefont {Zhang}}]{Zeng:2019yhi}%
  \BibitemOpen
  \bibfield  {author} {\bibinfo {author} {\bibfnamefont {H.-B.}\ \bibnamefont {Zeng}}, \bibinfo {author} {\bibfnamefont {C.-Y.}\ \bibnamefont {Xia}}, \ and\ \bibinfo {author} {\bibfnamefont {H.-Q.}\ \bibnamefont {Zhang}},\ }\bibfield  {title} {\enquote {\bibinfo {title} {Topological defects as relics of spontaneous symmetry breaking from black hole physics},}\ }\href {\doibase 10.1007/JHEP03(2021)136} {\bibfield  {journal} {\bibinfo  {journal} {Journal of High Energy Physics}\ }\textbf {\bibinfo {volume} {2021}},\ \bibinfo {pages} {136} (\bibinfo {year} {2021})}\BibitemShut {NoStop}%
\bibitem [{\citenamefont {del Campo}(2018)}]{delcampo18}%
  \BibitemOpen
  \bibfield  {author} {\bibinfo {author} {\bibfnamefont {A.}~\bibnamefont {del Campo}},\ }\bibfield  {title} {\enquote {\bibinfo {title} {Universal statistics of topological defects formed in a quantum phase transition},}\ }\href {\doibase 10.1103/PhysRevLett.121.200601} {\bibfield  {journal} {\bibinfo  {journal} {Phys. Rev. Lett.}\ }\textbf {\bibinfo {volume} {121}},\ \bibinfo {pages} {200601} (\bibinfo {year} {2018})}\BibitemShut {NoStop}%
\bibitem [{\citenamefont {G\'omez-Ruiz}\ \emph {et~al.}(2020)\citenamefont {G\'omez-Ruiz}, \citenamefont {Mayo},\ and\ \citenamefont {del Campo}}]{GomezRuiz20}%
  \BibitemOpen
  \bibfield  {author} {\bibinfo {author} {\bibfnamefont {F.~J.}\ \bibnamefont {G\'omez-Ruiz}}, \bibinfo {author} {\bibfnamefont {J.~J.}\ \bibnamefont {Mayo}}, \ and\ \bibinfo {author} {\bibfnamefont {A.}~\bibnamefont {del Campo}},\ }\bibfield  {title} {\enquote {\bibinfo {title} {Full counting statistics of topological defects after crossing a phase transition},}\ }\href {\doibase 10.1103/PhysRevLett.124.240602} {\bibfield  {journal} {\bibinfo  {journal} {Phys. Rev. Lett.}\ }\textbf {\bibinfo {volume} {124}},\ \bibinfo {pages} {240602} (\bibinfo {year} {2020})}\BibitemShut {NoStop}%
\bibitem [{\citenamefont {del Campo}\ \emph {et~al.}(2021)\citenamefont {del Campo}, \citenamefont {G{\'o}mez-Ruiz}, \citenamefont {Li}, \citenamefont {Xia}, \citenamefont {Zeng},\ and\ \citenamefont {Zhang}}]{delCampo21}%
  \BibitemOpen
  \bibfield  {author} {\bibinfo {author} {\bibfnamefont {A.}~\bibnamefont {del Campo}}, \bibinfo {author} {\bibfnamefont {F.~J.}\ \bibnamefont {G{\'o}mez-Ruiz}}, \bibinfo {author} {\bibfnamefont {Z.-H.}\ \bibnamefont {Li}}, \bibinfo {author} {\bibfnamefont {C.-Y.}\ \bibnamefont {Xia}}, \bibinfo {author} {\bibfnamefont {H.-B.}\ \bibnamefont {Zeng}}, \ and\ \bibinfo {author} {\bibfnamefont {H.-Q.}\ \bibnamefont {Zhang}},\ }\bibfield  {title} {\enquote {\bibinfo {title} {Universal statistics of vortices in a newborn holographic superconductor: beyond the {K}ibble-{Z}urek mechanism},}\ }\href {\doibase 10.1007/JHEP06(2021)061} {\bibfield  {journal} {\bibinfo  {journal} {Journal of High Energy Physics}\ }\textbf {\bibinfo {volume} {2021}},\ \bibinfo {pages} {61} (\bibinfo {year} {2021})}\BibitemShut {NoStop}%
\bibitem [{\citenamefont {del Campo}\ \emph {et~al.}(2022)\citenamefont {del Campo}, \citenamefont {G\'omez-Ruiz},\ and\ \citenamefont {Zhang}}]{delcampo22}%
  \BibitemOpen
  \bibfield  {author} {\bibinfo {author} {\bibfnamefont {A.}~\bibnamefont {del Campo}}, \bibinfo {author} {\bibfnamefont {F.~J.}\ \bibnamefont {G\'omez-Ruiz}}, \ and\ \bibinfo {author} {\bibfnamefont {H.-Q.}\ \bibnamefont {Zhang}},\ }\bibfield  {title} {\enquote {\bibinfo {title} {Locality of spontaneous symmetry breaking and universal spacing distribution of topological defects formed across a phase transition},}\ }\href {\doibase 10.1103/PhysRevB.106.L140101} {\bibfield  {journal} {\bibinfo  {journal} {Phys. Rev. B}\ }\textbf {\bibinfo {volume} {106}},\ \bibinfo {pages} {L140101} (\bibinfo {year} {2022})}\BibitemShut {NoStop}%
\bibitem [{\citenamefont {Nigmatullin}\ \emph {et~al.}(2016)\citenamefont {Nigmatullin}, \citenamefont {del Campo}, \citenamefont {De~Chiara}, \citenamefont {Morigi}, \citenamefont {Plenio},\ and\ \citenamefont {Retzker}}]{Nigmatullin16}%
  \BibitemOpen
  \bibfield  {author} {\bibinfo {author} {\bibfnamefont {R.}~\bibnamefont {Nigmatullin}}, \bibinfo {author} {\bibfnamefont {A.}~\bibnamefont {del Campo}}, \bibinfo {author} {\bibfnamefont {G.}~\bibnamefont {De~Chiara}}, \bibinfo {author} {\bibfnamefont {G.}~\bibnamefont {Morigi}}, \bibinfo {author} {\bibfnamefont {M.~B.}\ \bibnamefont {Plenio}}, \ and\ \bibinfo {author} {\bibfnamefont {A.}~\bibnamefont {Retzker}},\ }\bibfield  {title} {\enquote {\bibinfo {title} {Formation of helical ion chains},}\ }\href {\doibase 10.1103/PhysRevB.93.014106} {\bibfield  {journal} {\bibinfo  {journal} {Phys. Rev. B}\ }\textbf {\bibinfo {volume} {93}},\ \bibinfo {pages} {014106} (\bibinfo {year} {2016})}\BibitemShut {NoStop}%
\bibitem [{\citenamefont {Xia}\ and\ \citenamefont {Zeng}(2020)}]{Xia2020}%
  \BibitemOpen
  \bibfield  {author} {\bibinfo {author} {\bibfnamefont {C.-Y.}\ \bibnamefont {Xia}}\ and\ \bibinfo {author} {\bibfnamefont {H.-B.}\ \bibnamefont {Zeng}},\ }\bibfield  {title} {\enquote {\bibinfo {title} {Winding up a finite size holographic superconducting ring beyond kibble-zurek mechanism},}\ }\href {\doibase 10.1103/PhysRevD.102.126005} {\bibfield  {journal} {\bibinfo  {journal} {Phys. Rev. D}\ }\textbf {\bibinfo {volume} {102}},\ \bibinfo {pages} {126005} (\bibinfo {year} {2020})}\BibitemShut {NoStop}%
\bibitem [{\citenamefont {Lin}\ \emph {et~al.}(2014)\citenamefont {Lin}, \citenamefont {Wang}, \citenamefont {Kamiya}, \citenamefont {Chern}, \citenamefont {Fan}, \citenamefont {Fan}, \citenamefont {Casas}, \citenamefont {Liu}, \citenamefont {Kiryukhin}, \citenamefont {Zurek}, \citenamefont {Batista},\ and\ \citenamefont {Cheong}}]{Lin14}%
  \BibitemOpen
  \bibfield  {author} {\bibinfo {author} {\bibfnamefont {S.-Z.}\ \bibnamefont {Lin}}, \bibinfo {author} {\bibfnamefont {X.}~\bibnamefont {Wang}}, \bibinfo {author} {\bibfnamefont {Y.}~\bibnamefont {Kamiya}}, \bibinfo {author} {\bibfnamefont {G.-W.}\ \bibnamefont {Chern}}, \bibinfo {author} {\bibfnamefont {F.}~\bibnamefont {Fan}}, \bibinfo {author} {\bibfnamefont {D.}~\bibnamefont {Fan}}, \bibinfo {author} {\bibfnamefont {B.}~\bibnamefont {Casas}}, \bibinfo {author} {\bibfnamefont {Y.}~\bibnamefont {Liu}}, \bibinfo {author} {\bibfnamefont {V.}~\bibnamefont {Kiryukhin}}, \bibinfo {author} {\bibfnamefont {W.~H.}\ \bibnamefont {Zurek}}, \bibinfo {author} {\bibfnamefont {C.~D.}\ \bibnamefont {Batista}}, \ and\ \bibinfo {author} {\bibfnamefont {S.-W.}\ \bibnamefont {Cheong}},\ }\bibfield  {title} {\enquote {\bibinfo {title} {Topological defects as relics of emergent continuous symmetry and {H}iggs condensation of disorder in ferroelectrics},}\ }\href {\doibase 10.1038/nphys3142} {\bibfield  {journal} {\bibinfo
  {journal} {Nature Physics}\ }\textbf {\bibinfo {volume} {10}},\ \bibinfo {pages} {970} (\bibinfo {year} {2014})}\BibitemShut {NoStop}%
\bibitem [{\citenamefont {Gardiner}\ and\ \citenamefont {Davis}(2003)}]{Gardiner_2003}%
  \BibitemOpen
  \bibfield  {author} {\bibinfo {author} {\bibfnamefont {C.~W.}\ \bibnamefont {Gardiner}}\ and\ \bibinfo {author} {\bibfnamefont {M.~J.}\ \bibnamefont {Davis}},\ }\bibfield  {title} {\enquote {\bibinfo {title} {The stochastic gross–pitaevskii equation: Ii},}\ }\href {\doibase 10.1088/0953-4075/36/23/010} {\bibfield  {journal} {\bibinfo  {journal} {Journal of Physics B: Atomic, Molecular and Optical Physics}\ }\textbf {\bibinfo {volume} {36}},\ \bibinfo {pages} {4731–4753} (\bibinfo {year} {2003})}\BibitemShut {NoStop}%
\bibitem [{\citenamefont {{P. B. Blakie, A. S. Bradley, M. J. Davis, R. J. Ballagh and C. W. Gardiner}}(2008)}]{blakie2008dynamics}%
  \BibitemOpen
  \bibfield  {author} {\bibinfo {author} {\bibnamefont {{P. B. Blakie, A. S. Bradley, M. J. Davis, R. J. Ballagh and C. W. Gardiner}}},\ }\bibfield  {title} {\enquote {\bibinfo {title} {{Dynamics and statistical mechanics of ultra-cold Bose gases using c-field techniques}},}\ }\href {\doibase 10.1080/00018730802564254} {\bibfield  {journal} {\bibinfo  {journal} {Advances in Physics}\ }\textbf {\bibinfo {volume} {57}},\ \bibinfo {pages} {363} (\bibinfo {year} {2008})}\BibitemShut {NoStop}%
\bibitem [{\citenamefont {Rooney}\ \emph {et~al.}(2012)\citenamefont {Rooney}, \citenamefont {Blakie},\ and\ \citenamefont {Bradley}}]{Rooney2012}%
  \BibitemOpen
  \bibfield  {author} {\bibinfo {author} {\bibfnamefont {S.~J.}\ \bibnamefont {Rooney}}, \bibinfo {author} {\bibfnamefont {P.~B.}\ \bibnamefont {Blakie}}, \ and\ \bibinfo {author} {\bibfnamefont {A.~S.}\ \bibnamefont {Bradley}},\ }\bibfield  {title} {\enquote {\bibinfo {title} {Stochastic projected gross-pitaevskii equation},}\ }\href {\doibase 10.1103/PhysRevA.86.053634} {\bibfield  {journal} {\bibinfo  {journal} {Phys. Rev. A}\ }\textbf {\bibinfo {volume} {86}},\ \bibinfo {pages} {053634} (\bibinfo {year} {2012})}\BibitemShut {NoStop}%
\bibitem [{\citenamefont {Rooney}\ \emph {et~al.}(2014)\citenamefont {Rooney}, \citenamefont {Blakie},\ and\ \citenamefont {Bradley}}]{Rooney2014}%
  \BibitemOpen
  \bibfield  {author} {\bibinfo {author} {\bibfnamefont {S.~J.}\ \bibnamefont {Rooney}}, \bibinfo {author} {\bibfnamefont {P.~B.}\ \bibnamefont {Blakie}}, \ and\ \bibinfo {author} {\bibfnamefont {A.~S.}\ \bibnamefont {Bradley}},\ }\bibfield  {title} {\enquote {\bibinfo {title} {Numerical method for the stochastic projected gross-pitaevskii equation},}\ }\href {\doibase 10.1103/PhysRevE.89.013302} {\bibfield  {journal} {\bibinfo  {journal} {Phys. Rev. E}\ }\textbf {\bibinfo {volume} {89}},\ \bibinfo {pages} {013302} (\bibinfo {year} {2014})}\BibitemShut {NoStop}%
\bibitem [{\citenamefont {Bradley}\ \emph {et~al.}(2015)\citenamefont {Bradley}, \citenamefont {Rooney},\ and\ \citenamefont {McDonald}}]{Bradley2015}%
  \BibitemOpen
  \bibfield  {author} {\bibinfo {author} {\bibfnamefont {A.~S.}\ \bibnamefont {Bradley}}, \bibinfo {author} {\bibfnamefont {S.~J.}\ \bibnamefont {Rooney}}, \ and\ \bibinfo {author} {\bibfnamefont {R.~G.}\ \bibnamefont {McDonald}},\ }\bibfield  {title} {\enquote {\bibinfo {title} {Low-dimensional stochastic projected gross-pitaevskii equation},}\ }\href {\doibase 10.1103/PhysRevA.92.033631} {\bibfield  {journal} {\bibinfo  {journal} {Phys. Rev. A}\ }\textbf {\bibinfo {volume} {92}},\ \bibinfo {pages} {033631} (\bibinfo {year} {2015})}\BibitemShut {NoStop}%
\bibitem [{\citenamefont {McDonald}\ \emph {et~al.}(2020)\citenamefont {McDonald}, \citenamefont {Barnett}, \citenamefont {Atayee},\ and\ \citenamefont {Bradley}}]{McDonald2020}%
  \BibitemOpen
  \bibfield  {author} {\bibinfo {author} {\bibfnamefont {R.~G.}\ \bibnamefont {McDonald}}, \bibinfo {author} {\bibfnamefont {P.~S.}\ \bibnamefont {Barnett}}, \bibinfo {author} {\bibfnamefont {F.}~\bibnamefont {Atayee}}, \ and\ \bibinfo {author} {\bibfnamefont {A.~S.}\ \bibnamefont {Bradley}},\ }\bibfield  {title} {\enquote {\bibinfo {title} {{Dynamics of hot Bose-Einstein condensates: stochastic Ehrenfest relations for number and energy damping}},}\ }\href {\doibase 10.21468/SciPostPhys.8.2.029} {\bibfield  {journal} {\bibinfo  {journal} {SciPost Phys.}\ }\textbf {\bibinfo {volume} {8}},\ \bibinfo {pages} {029} (\bibinfo {year} {2020})}\BibitemShut {NoStop}%
\bibitem [{\citenamefont {Laguna}\ and\ \citenamefont {Zurek}(1997)}]{Laguna97}%
  \BibitemOpen
  \bibfield  {author} {\bibinfo {author} {\bibfnamefont {P.}~\bibnamefont {Laguna}}\ and\ \bibinfo {author} {\bibfnamefont {W.~H.}\ \bibnamefont {Zurek}},\ }\bibfield  {title} {\enquote {\bibinfo {title} {Density of kinks after a quench: When symmetry breaks, how big are the pieces?}}\ }\href {\doibase 10.1103/PhysRevLett.78.2519} {\bibfield  {journal} {\bibinfo  {journal} {Phys. Rev. Lett.}\ }\textbf {\bibinfo {volume} {78}},\ \bibinfo {pages} {2519} (\bibinfo {year} {1997})}\BibitemShut {NoStop}%
\bibitem [{\citenamefont {Yates}\ and\ \citenamefont {Zurek}(1998)}]{Yates98}%
  \BibitemOpen
  \bibfield  {author} {\bibinfo {author} {\bibfnamefont {A.}~\bibnamefont {Yates}}\ and\ \bibinfo {author} {\bibfnamefont {W.~H.}\ \bibnamefont {Zurek}},\ }\bibfield  {title} {\enquote {\bibinfo {title} {Vortex formation in two dimensions: When symmetry breaks, how big are the pieces?}}\ }\href {\doibase 10.1103/PhysRevLett.80.5477} {\bibfield  {journal} {\bibinfo  {journal} {Phys. Rev. Lett.}\ }\textbf {\bibinfo {volume} {80}},\ \bibinfo {pages} {5477} (\bibinfo {year} {1998})}\BibitemShut {NoStop}%
\bibitem [{\citenamefont {Balducci}\ \emph {et~al.}(2023)\citenamefont {Balducci}, \citenamefont {Beau}, \citenamefont {Yang}, \citenamefont {Gambassi},\ and\ \citenamefont {del Campo}}]{Balducci23}%
  \BibitemOpen
  \bibfield  {author} {\bibinfo {author} {\bibfnamefont {F.}~\bibnamefont {Balducci}}, \bibinfo {author} {\bibfnamefont {M.}~\bibnamefont {Beau}}, \bibinfo {author} {\bibfnamefont {J.}~\bibnamefont {Yang}}, \bibinfo {author} {\bibfnamefont {A.}~\bibnamefont {Gambassi}}, \ and\ \bibinfo {author} {\bibfnamefont {A.}~\bibnamefont {del Campo}},\ }\bibfield  {title} {\enquote {\bibinfo {title} {Large deviations beyond the kibble-zurek mechanism},}\ }\href {\doibase 10.1103/PhysRevLett.131.230401} {\bibfield  {journal} {\bibinfo  {journal} {Phys. Rev. Lett.}\ }\textbf {\bibinfo {volume} {131}},\ \bibinfo {pages} {230401} (\bibinfo {year} {2023})}\BibitemShut {NoStop}%
\bibitem [{\citenamefont {url will be inserted~by publisher}()}]{SM}%
  \BibitemOpen
  \bibfield  {author} {\bibinfo {author} {\bibnamefont {url will be inserted~by publisher}},\ }\href@noop {} {\ }\BibitemShut {NoStop}%
\bibitem [{\citenamefont {Bradley}\ and\ \citenamefont {Anderson}(2012)}]{Bradley12}%
  \BibitemOpen
  \bibfield  {author} {\bibinfo {author} {\bibfnamefont {A.~S.}\ \bibnamefont {Bradley}}\ and\ \bibinfo {author} {\bibfnamefont {B.~P.}\ \bibnamefont {Anderson}},\ }\bibfield  {title} {\enquote {\bibinfo {title} {Energy spectra of vortex distributions in two-dimensional quantum turbulence},}\ }\href {\doibase 10.1103/PhysRevX.2.041001} {\bibfield  {journal} {\bibinfo  {journal} {Phys. Rev. X}\ }\textbf {\bibinfo {volume} {2}},\ \bibinfo {pages} {041001} (\bibinfo {year} {2012})}\BibitemShut {NoStop}%
\bibitem [{\citenamefont {Chiu}\ \emph {et~al.}(2013)\citenamefont {Chiu}, \citenamefont {Stoyan}, \citenamefont {Kendall},\ and\ \citenamefont {Mecke}}]{StochasticGeometry2013}%
  \BibitemOpen
  \bibfield  {author} {\bibinfo {author} {\bibfnamefont {S.~N.}\ \bibnamefont {Chiu}}, \bibinfo {author} {\bibfnamefont {D.}~\bibnamefont {Stoyan}}, \bibinfo {author} {\bibfnamefont {W.~S.}\ \bibnamefont {Kendall}}, \ and\ \bibinfo {author} {\bibfnamefont {J.}~\bibnamefont {Mecke}},\ }\enquote {\bibinfo {title} {Line, fibre and surface processes},}\ in\ \href {\doibase https://doi.org/10.1002/9781118658222.ch08} {\emph {\bibinfo {booktitle} {Stochastic Geometry and its Applications}}}\ (\bibinfo  {publisher} {John Wiley \& Sons, Ltd, Chichester},\ \bibinfo {year} {2013})\ Chap.~\bibinfo {chapter} {8}, pp.\ \bibinfo {pages} {297--342}\BibitemShut {NoStop}%
\end{thebibliography}%

\newpage
\clearpage

\title{---Supplementary Material--- \\Circulation Statistics and Migdal Area Rule Beyond the Kibble-Zurek Mechanism in a Newborn Bose-Einstein Condensate
}
\maketitle
\onecolumngrid

\renewcommand{\theequation}{S\arabic{equation}}
\renewcommand{\thefigure}{S\arabic{figure}}
\renewcommand{\thetable}{S\arabic{table}}

\setcounter{equation}{0}
\setcounter{figure}{0}
\setcounter{table}{0}
\setcounter{page}{1}

\section{Calculation of the moments of the net winding number\\ trapped inside a loop $C$}
We 
derive explicit expressions for the moments of the net winding number $W$ enclosed by a loop $C$ with contour length $|C|$, delimiting an area of size $|A|$. 
Starting from the moment generating function 
$
    \langle e^{Wt} \rangle=\prod_{i=1}^{\mathcal{N}} \langle e^{z_{i}t} \rangle=\left[1+\lambda\left(\cosh(t)-1\right)\right]^{\mathcal{N}}
$, 
the first few even moments are:
\begin{align}\label{moments_of_W}
    &\langle W^{2} \rangle = \lambda \mathcal{N},\nonumber\\
    &\langle W^{4} \rangle =\lambda \mathcal{N}+ 3\lambda^{2} \left(\mathcal{N}^{2}-\mathcal{N}\right), \nonumber\\
    & \langle W^{6} \rangle=\lambda \mathcal{N}+15 \lambda^{2}\left(\mathcal{N}^{2}-\mathcal{N}\right)+15 \lambda^{3}\left(\mathcal{N}^{3}-3\mathcal{N}^{2}+2\mathcal{N}\right), \nonumber\\
    &\langle W^{8} \rangle= \lambda \mathcal{N}+63 \lambda^{2} \left(\mathcal{N}^{2}-\mathcal{N}\right)+210\lambda^{3}\left(\mathcal{N}^{3}-3\mathcal{N}^{2}+2\mathcal{N}\right)+105\lambda^{4}\left(\mathcal{N}^{4}-6\mathcal{N}^{3}+11 \mathcal{N}^{2}-6\mathcal{N}\right),
\end{align}
where $\lambda:= p_{v}p_{C}$, as defined in the main text. 
All odd moments vanish due to the symmetry of $P(W)$ where 
\begin{equation}
    P(W=n)=\sum_{\substack{m = |n| \\ m + n \, \text{even}}}^{\mathcal{N}}  \binom{\mathcal{N}}{m}\binom{m}{\frac{m+n}{2}}\left(\frac{\lambda}{2}\right)^{m}(1-\lambda)^{\mathcal{N}-m}.
\end{equation}
We now focus on the two asymptotic regimes for the loop $C$.

In the small loop limit, \( p_{C} \to 0 \), so the leading contribution to each even moment is $\lambda \mathcal{N}$. Noting that $p_{C}=2|A|/A_{tot}$, and $\mathcal{N}=A_{tot}/(2\hat{\xi}^{2}f)$ one finds
\begin{equation}\label{W^2p_small_A}
    \langle W^{2p} \rangle = p_v p_{C} \mathcal{N}=\frac{p_{v}}{\hat{\xi}^{2}f} |A|, \quad \text{for} \quad |A| \ll \hat{\xi}^{2}.
\end{equation}

In the opposite limit, $|A| \gg \hat{\xi}^{2}$, we have
\begin{equation}\label{W^2p}
    \langle W^{2p} \rangle =(2p-1)!! (\mathcal{N} \lambda)^{p} =(2p-1)!! \left(\frac{p_{v}|C|}{\pi \hat{\xi}f}\right)^{p},
\end{equation}
where the last equality follows from $p_{C}=2\hat{\xi}|C|/(\pi A_{tot})$, valid in the large loop regime. 
The expression for $p_{C}$ can be derived from the Poisson line process \cite{StochasticGeometry2013} by noting that the total number of lines is $\mathcal{N} p_v$.

\begin{figure}[h]
    \centering
    \includegraphics[width=0.5\linewidth]{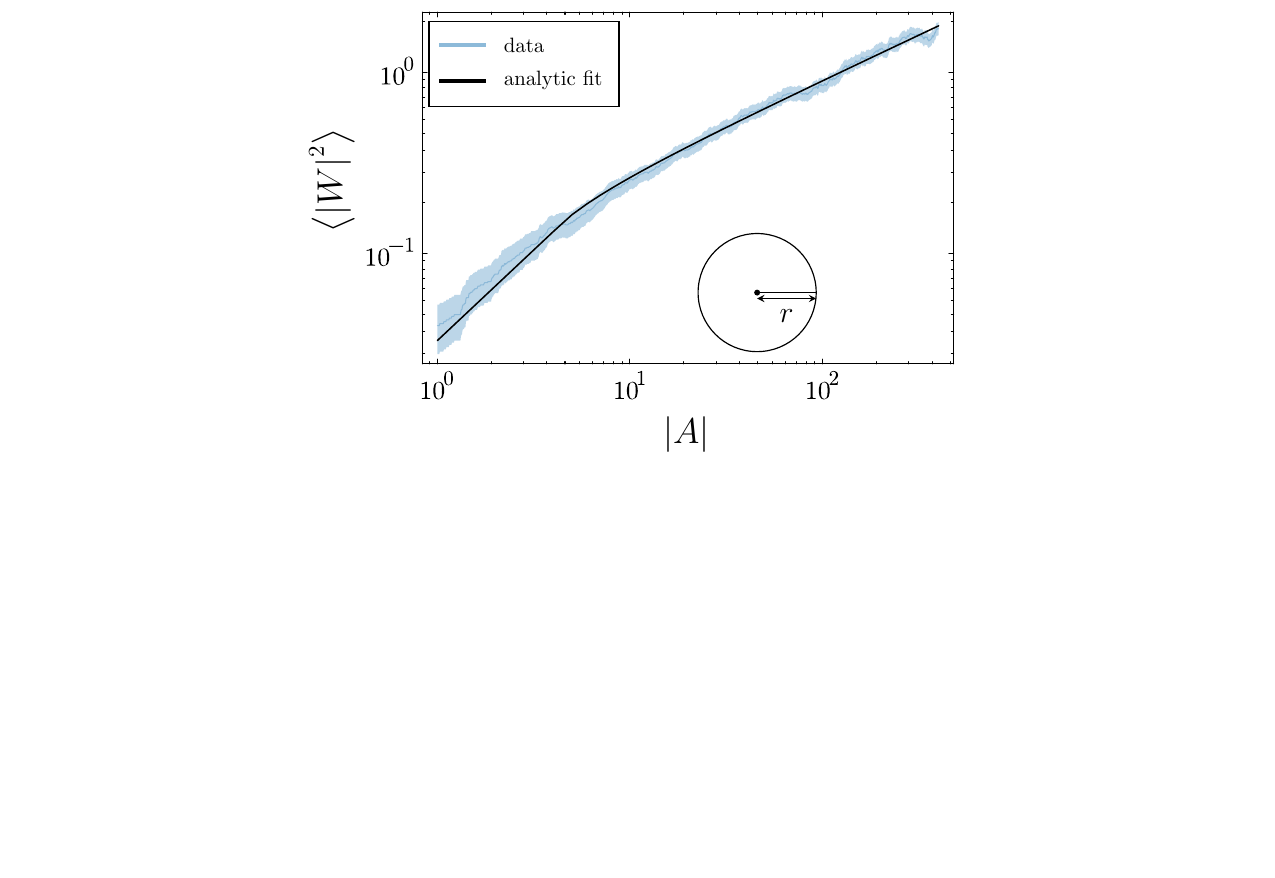}
    \caption{Second moment of the net change within a circular loop as a function of its area. The blue curve corresponds to the data at equilibration time after a quench of $\tau_{Q}=50$, averaged over $\mathcal{R}=1000$ noise realizations, with the shaded band indicating $95\%$ confidence interval. The black curve depicts the analytic expression for the second moment based on Eqs. (\ref{moments_of_W}) and (\ref{pc_expression_circle}), where $\hat{\xi}$ and $p_v \mathcal{N}$ are the fitting parameters used, with fitted values $\hat{\xi}=2.43 \pm 0.11$ and $p_v \mathcal{N}=14.86 \pm 0.64$.}
    \label{fig:W2_vs_A_data_vs_analytic_fit}
\end{figure}

We now carry out the same analysis for the moments of $|W|$.
In particular, for a small loop, $P(W)$
asymptotically approaches
\begin{equation}
    P(W=n)\approx \binom{\mathcal{N}}{|n|} \left( \frac{\lambda}{2}\right)^{|n|}\left[1-(\mathcal{N}-|n| )\lambda\right]. 
\end{equation}
and the $p$-th moment behaves as
\begin{equation}\label{|W|^p_small_A}
    \langle |W|^{p} \rangle= \lambda\mathcal{N},
\end{equation}
in agreement with Eq. (\ref{W^2p_small_A}).
In the big loop scenario, we can invoke the central limit theorem, which yields 
\begin{equation}
    P(W)\approx \frac{1}{\sqrt{2\pi \lambda \mathcal{N}}} \exp\left(-\frac{W^2}{2 \lambda \mathcal{N}}\right),
\end{equation}
therefore
\begin{equation}\label{|W|^p}
    \langle |W|^{p} \rangle=\frac{2^{p/2}}{\sqrt{\pi}}\Gamma\left(\frac{1+p}{2}\right)(\lambda \mathcal{N})^{p / 2}=\frac{2^{p}}{\sqrt{\pi}}\Gamma\left(\frac{1+p}{2}\right)\left(\frac{p_{v}|C|}{2\pi \hat{\xi}f}\right)^{p / 2}.
\end{equation}
Additionally, we note that, although we have focused on the two asymptotic limits, Eqs. (\ref{moments_of_W}) in fact provides the full interpolating behavior of $\langle W^{2p} \rangle$ between the small and large loop regimes. In particular, for a circular loop of radius $r$, the crossing probability $p_{C}$ takes the form
\begin{equation}\label{pc_expression_circle}
p_{C}=
\begin{cases}
2\frac{\pi r^2}{A_{tot}}, \quad  \quad r<\frac{\hat{\xi}}{2}\\
\\
\frac{2}{A_{\text{tot}}} \left[\frac{1}{2} \hat{\xi} \sqrt{4r^2 - \hat{\xi}^2} - r^2 \left( 2 \cos^{-1} \left( \frac{\hat{\xi}}{2r} \right) - \pi \right) \right], \quad  \quad r>\frac{\hat{\xi}}{2}.
\end{cases}
\end{equation}
As an example, in Fig. \ref{fig:W2_vs_A_data_vs_analytic_fit} we plot the full behavior of $\langle W^{2} \rangle$ for a circular loop as a function of its area.

Another way of expressing Eq. (\ref{|W|^p_small_A}), which directly follows from the fact that $\langle n\rangle=2\mathcal{N}p_{v}$ is the KZ average vortex number, is
\begin{equation}
    \langle |W|^{p}\rangle=\rho_{\rm KZ}|A|=\langle n_{A} \rangle.
\end{equation}
This could also have been deduced from the PPP model with the additional assumption that, for small loops, at most one vortex can be found inside it \cite{Zurek13}. However, within the vortex pair model, the result naturally comes as a geometric consequence. 
Similarly, we can express Eq. (\ref{|W|^p}) in terms of the average number of vortices trapped inside the loop $\langle n_{A}\rangle$ as
\begin{equation}\label{|W|^{p}_vs_nA_big_loop}
    \langle |W|^{p} \rangle=\frac{2^{p}}{\sqrt{\pi}}\Gamma\left(\frac{p+1}{2}\right)\left(\frac{k}{2\pi}\sqrt{\frac{p_{v}}{f}} \sqrt{\langle n_{A}\rangle}\right)^{p / 2},
\end{equation}
for the case where the length of the loop $|C|$ and the area $|A|$ enclosed are related via $|C|=k \sqrt{|A|}$, with $k$ a shape-dependent constant.

\begin{figure}[hptb]
\includegraphics[width=0.7\columnwidth]{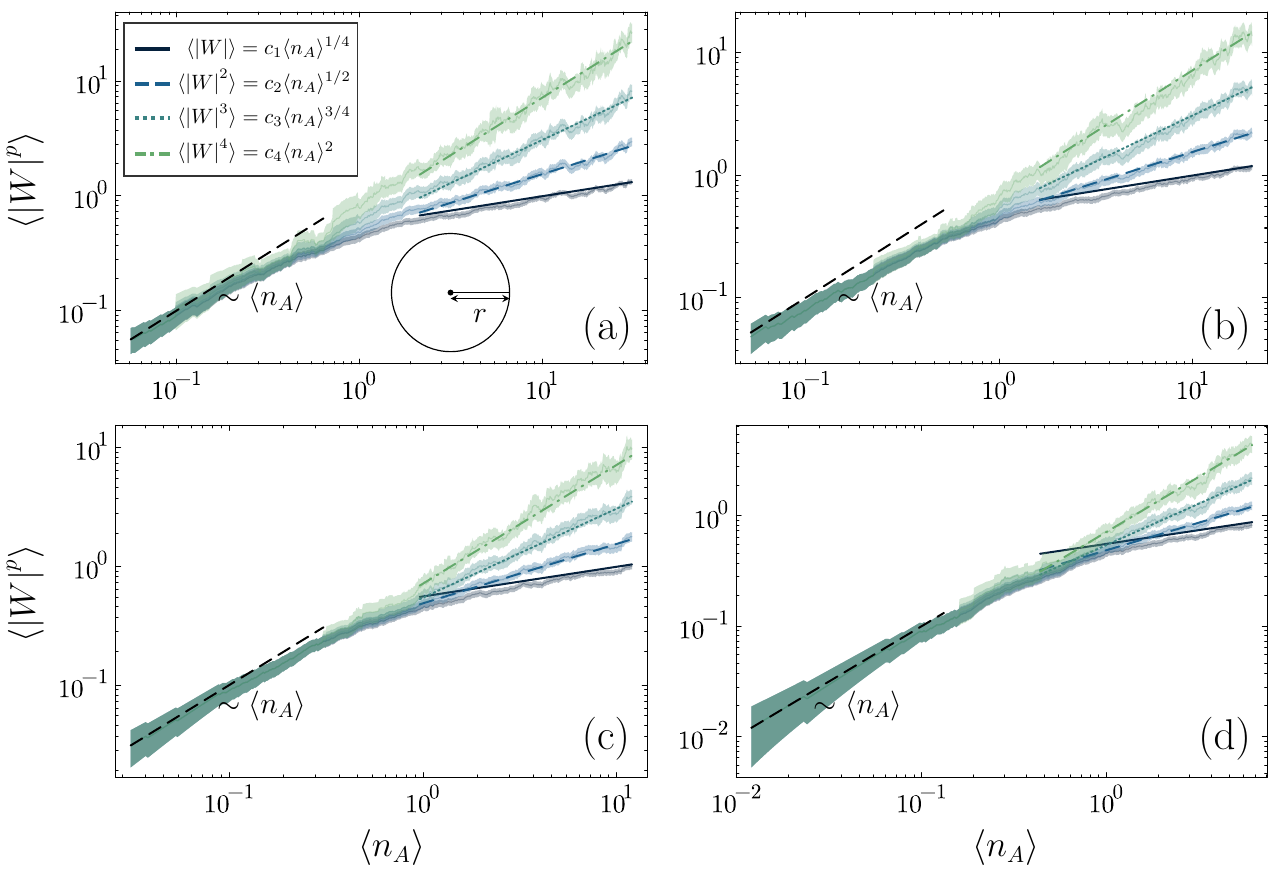}
\caption{\textbf{Moments of the absolute net vortex charge \(|W|\) in a circular contour versus the mean vortex number \(\langle n_A\rangle\).}  
Panels (a)–(d) show the \(p\)-th moments of $\langle |W| \rangle $ (\(p=1,2,3,4\)) at quench rates \(\tau_Q=10,30,70,200\), respectively, each averaged over \(\mathcal{R}=1000\) noise realizations. The fitting lines in the large \(\langle n_A\rangle\) regime correspond to Eq.~(\ref{|W|^{p}_vs_nA_big_loop}) using \(p_v/(f\pi)= 0.0596\), as determined from the 
fourth-moment 
data for \(\tau_Q=10\). This yields the prefactors \(c_1=0.558\), 
\(c_2=0.488\), \(c_3=0.544\), and \(c_4=0.715\). 
Shaded bands denote 95\% confidence intervals.
}
\label{fig:W_vs_nv}
\end{figure}

The validity of the vortex pair model is tested for the circular area in Fig. \ref{fig:W_vs_nv}, with the value 
$p_v / f \pi = 0.0596 \pm 0.0003$ found from the fourth moment of the net charge 
data at $\tau_Q = 10$, showing that our model provides an accurate description for both small and large loop regions. 

\section{Additional plots}

\begin{figure}[h!]
\includegraphics[width=0.8\columnwidth]{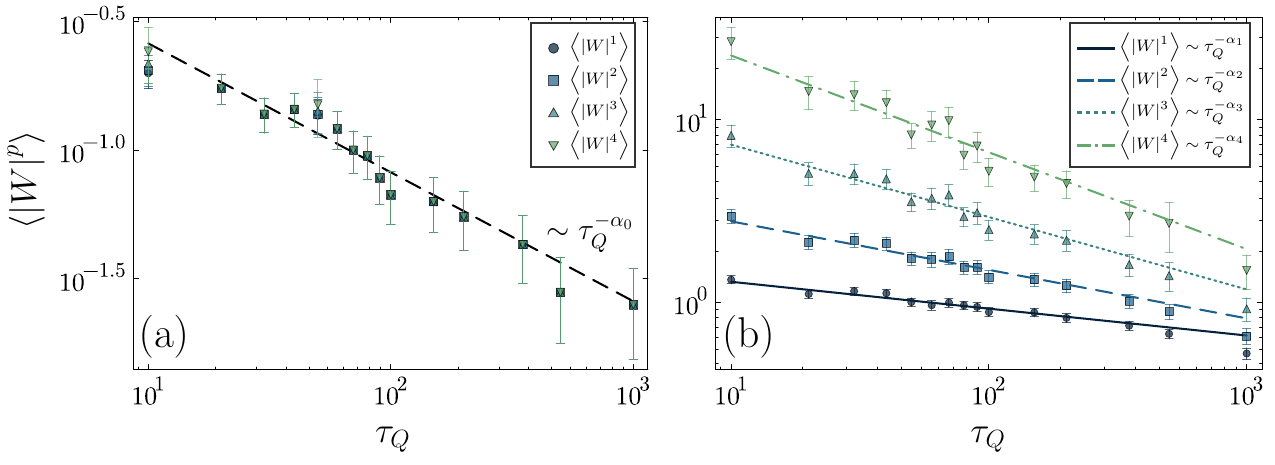}
\caption{\textbf{Moments of the absolute net vortex charge \(|W|\) within a fixed circular area \(A\) versus quench time \(\tau_Q\).} Panel (a) shows the small-area case (\(|A|=4\)) 
where $\alpha_0 = 0.502 \pm 0.040$, which is close to $1/2$ expected from the vortex pair model, and panel (b) the large-area case (\(|A|=400\)) where $\alpha_1 = 0.147 \pm 0.018$, 
$\alpha_2 = 0.266 \pm 0.031$, 
$\alpha_3 = 0.396 \pm 0.048$, 
and $\alpha_4 = 0.530 \pm 0.067$, which are close to $\alpha_j = j / 8$ expected from the vortex pair model. Data are averaged over \(\mathcal{R}=1000\) independent noise realizations, with error bars indicating 95\% confidence intervals.
}
\label{fig:W_vs_Tq_small_A}
\end{figure}

We present the moments of the absolute net vortex charge within a fixed circular area versus quench time in Figs. \ref{fig:W_vs_Tq_small_A}. 


\section{Details on the different loop geometries}
The $\mathbb{C}$-shaped loop used to test the area rule is a semicircular annulus of inner radius $r_{1}$ and outer radius $b r_{1}$. The ratio $b$ was chosen to highlight the difference with rectangular and circular contours. In particular, for each loop we denote its enclosed area by $|A|$ and its contour length by $|C|$, related by $|C|=k\sqrt{|A|}$ with shape-dependent constant $k$. 
For the circle, $k_{circ} = 2 \sqrt{\pi} \simeq 3.545$. For a rectangle with aspect ratio $a$, $k_{rect} = 2 \left( 1 + a \right) / \sqrt{a}$, and for the $\mathbb{C}$-shaped loop  $k_{\mathbb{C}} = \left\lbrack \left( \pi + 2 \right) b + \pi - 2 \right\rbrack \sqrt{2 / \pi \left( b^2 - 1 \right)}$. 
In the large-loop limit, the circulation statistics depend on $|C|$, and plotting the circulation moments as a function of $|A|$ reveals the shape dependence at large $|A|$, encoded in the constant $k$. To emphasize the breakdown of universal area scaling, we selected loop geometries whose $k$ values differ substantially. 
Specifically, in the main text, we set $a = 4$ and $b = 1.46$, giving $k_{rect} = 5$ and $k_{\mathbb{C}} \simeq 6.487$.

\end{document}